\documentclass[11pt]{article}

\usepackage{amssymb}
\usepackage{amsmath}
\usepackage{graphicx}
\usepackage{fancyhdr}
\usepackage[left=1in,top=1in,right=1in,bottom=1in]{geometry} 
\usepackage{wasysym}
\usepackage{setspace}
\usepackage{xcolor}
\usepackage{framed}

\def\Carbon#1{^{#1}\text C}
\def\delC{\delta^{13}\text C}

\fancypagestyle{firststyle}
{
   \fancyhf{}
   \fancyfoot[C]{\thepage}
}

\begin{document}

\title{A new method of carbon budget analysis with application to present and paleo atmospheric concentration data sets}

  
  \author{
Alice Nadeau$^{1}$, Richard McGehee$^{2}$ and Clarence Lehman$^{3}$}

\date{%
\small
    $^{1}$Department of Mathematics, Cornell University \\%
    $^{2}$School of Mathematics, University of Minnesota\\%
    $^{3}$Department of Ecology, Evolution, and Behavior, University of Minnesota
}

  
%
%

\maketitle

\thispagestyle{firststyle}

\begin{abstract}
We introduce a new method for analyzing the carbon budget using box models and a mass balance approach. The method describes the net flow of carbon between the atmosphere and other reservoirs. The method  assumes that the data can be explained by having the carbon move in and out of the atmosphere through two reservoirs, one consisting of isotopically light carbon (biotic) and the other consisting of isotopically heavy carbon (abiotic). The systems are underdetermined from the data, so the Occam's razor approach is to assume that the least amount of carbon moves between the atmosphere and the reservoirs. As data from known sources and sinks are added to the model, one can determine the constraints on the unknown sources and sinks. To illustrate the method, we analyze data from the Mauna Loa Observatory and data from Antarctic ice cores.
\end{abstract}


\section{Introduction}
Earth's carbon cycle is highly complex, involving geophysical, geochemical, and biological processes occurring both on land and in the oceans.  The atmosphere is central to the cycle, since it communicates directly with the ocean and the land.  Carbon in the atmosphere and shallow ocean (which is in equilibrium with the atmosphere over short time scales) is converted to biological material through photosynthesis, and carbon is returned to the atmosphere and oceans through respiration by animals and micro-organisms.  On longer  time scales, carbon is transported to and from the deep ocean and to and from terrestrial reservoirs such as soils, lake sediments, and peatlands.  On an even longer time scale, carbon is deposited on the ocean floor, subducted via plate tectonics, and released back into the atmosphere by volcanic activity.

We can think of the various sources and sinks of atmospheric carbon as reservoirs.  Our goal here is to use  atmospheric carbon data to put constraints on the aggregate behavior of these reservoirs.

There are two stable isotopes of carbon: carbon-12 (the more abundant isotope, designated $\Carbon{12}$) and carbon-13 (the heavier and less abundant isotope, designated $\Carbon{13}$).  Plants take carbon dioxide and water from their environments and produce carbohydrates through photosynthesis.  The process discriminates in favor of the lighter $\Carbon{12}$ isotope, leaving the environment with a higher concentration of the heavier carbon \cite{Farquhar}.  All of the other organic compounds produced by the plants use the carbon captured through photosynthesis, and hence also have a lower proportion of $^{12}$C than the environment.  Animals get their supply of carbon by consuming plants and consequently also have the same lower proportion.

Roughly 1\% of the carbon on Earth is $\Carbon{13}$, while about 99\% is $\Carbon{12}$.  The ratio $R$ of $\Carbon{13}$ to $\Carbon{12}$ in a particular sample is not usually presented as a ratio, but is instead compared to a standard ratio $R_R$ using the formula
\begin{equation}                                  \label{eq:delC}
  \delC = 1000\times\left( \frac{R}{R_R} - 1 \right).
\end{equation}
Multiplication by $1000$ means that the value is expressed as ``per mil" ($\permil$).  For example, if the ratio $R$ in the sample is 1\% lower than the ratio $R_R$ of the standard sample, i.e., $R=0.99R_R$, then the result would be expressed $\delC = -10\permil$.

The timescale for photosynthesis and respiration is very rapid compared with the timescale for geological processes.  However, the biological processes can produce net gains and losses of atmospheric carbon occurring over centuries and millennia (e.g. \cite{Gorham12}).  Here we simply classify long-term sinks and sources as either \textit{biotic}, referring to a biological origin of the carbon, or \textit{abiotic}, referring to a non-biological origin of the carbon.  We then determine constraints on these reservoirs determined by the atmospheric carbon data.

\begin{figure}
\begin{framed}
\begin{center}
\includegraphics[width=.75\textwidth]{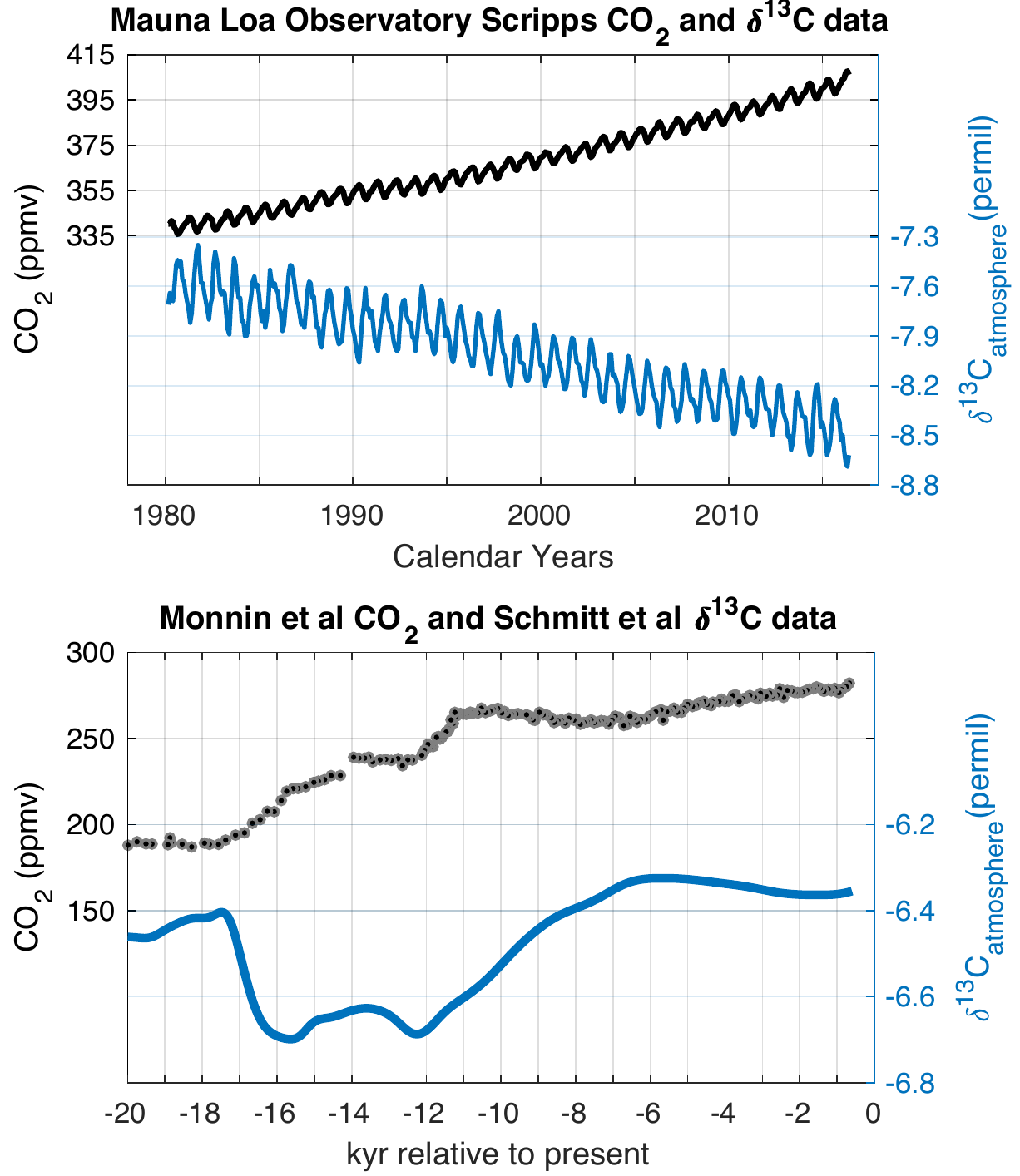}
\caption{Data used in this study. Top: Time series for total atmospheric carbon and $\delC$ since 1980. Data gathered by Scripps Carbon Dioxide Program at Mauna Loa Observatory in Hawaii \cite{Keeling}. Bottom: Time series for total atmospheric carbon and $\delC$ for the past 20,000 years. Total carbon dioxide collected by Monnin~\emph{et~al}~\cite{Monnin} and $\delC$ gathered and smoothed by Schmitt ~\emph{et~al}~ \cite{Schmitt}.}
\label{fig:data}
\end{center}
\end{framed}
\end{figure}

Below we describe the proposed method by building up the complexity of the model through several conceptual methods.  These methods each determine constraints on the nature of the carbon sinks and sources with varying assumptions. We illustrate the benefits and disadvantages of each method using two data sets.   The first data set comes from the Scripps Carbon Dioxide Program collected at the Mauna Loa Observatory \cite{Keeling} (replotted in Figure \ref{fig:data}, top).  Here our methods find a large source of biotic carbon consistent with the well-established conclusion that the dramatic recent increase in atmospheric carbon comes from the burning of fossil fuels.  When we add the fossil fuel data to the model, a somewhat mysterious source of abiotic carbon appears necessary to explain the data.  The second data set comes from Antarctic ice core records from the past 20,000 years given by Schmitt~\emph{et~al}~\cite{Schmitt}  and Monnin~\emph{et al} \cite{Monnin} (replotted in Figure \ref{fig:data}, bottom).  We find that our method yields constraints on the carbon budget consistent with  the conclusions in the Schmitt paper.  We further illustrate the method by adding data from North American peatlands given by Gorham~\emph{et~al}~\cite{Gorham12}.  There we find constraints consistent with the Ruddiman hypothesis \cite{Ruddiman, Kaplan} that early human agriculture has played a role in Earth's carbon cycle. 

We describe our data analysis methods by working through these examples. In Section~\ref{sect:MLO-sect} and Section~\ref{sect:paleo-sect}, we describe the methods using the Mauna Loa data and paleocarbon data, respectively. In Section~\ref{sect:methods} we provide a discussion of the methods, and in Section~\ref{sect:conclusion} we give our concluding remarks.

\section{Current Carbon Data} \label{sect:MLO-sect}

The Scripps Carbon Dioxide Program has been collecting samples of atmospheric carbon since the late 1950's at several stations within the Pacific basin \cite{Keeling}.  The most famous of these is the Mauna Loa Observatory (MLO) located on Mauna Loa in Hawaii.  At this station, they have been measuring the parts per million of carbon dioxide in the atmosphere on a monthly basis since 1958 and the isotopic $^{13}$C ratio monthly since 1980 \cite{Keeling}.

Over this period the concentration of atmospheric carbon increased, while the isotopic composition decreased as shown in Figure~\ref{fig:data}.  In addition to these general trends, the monthly records also show strong seasonal cycles due to the growth and death of the northern hemisphere biosphere over the course of a year \cite{Keeling}.

\subsection{Transforming the data} \label{sect:transforming}
The Scripps data that we use are reported as a monthly time series of concentrations of atmospheric carbon dioxide, given in parts per million by volume (ppmv), and $\delC$, as described above.  For our analysis, we wish to keep track of the total amounts of carbon-12 and carbon-13 in the atmosphere, which we will quantify in terms of petagrams.  A petagram (Pg) is $10^{15}$ grams, also often called a a gigatonne, or $10^9$ metric tonnes.  A metric tonne is a thousand kilograms, or a million grams.  Since the total mass of the atmosphere is known, we can convert the ppmv value in the data to total mass of carbon in the atmosphere via the ratio 2.13 Pg C/ppmv , arriving at a time series of the total petagrams of atmospheric carbon, which we will denote $C(t)$, where $t$ is time.

The $\delC(t)$ values in the Scripps data use the ``VPDB'' standard value of $R_R = 0.0112372$ \cite{VPDB} for the carbon-13 to carbon-12 ratio in equation~\eqref{eq:delC}.  Therefore, the isotope ratio $R$ we use in our analysis is
\begin{equation}                                   \label{eq:R}                                               
  R(t) = R_R\cdot(1 + 0.001\cdot\delC(t)),
\end{equation}
where $\delC(t)$ is the value appearing in the Scripps time series.

Now that we have the ratio $R(t)$ of carbon-13 to carbon-12 and the total mass $C(t)$ of atmospheric carbon, we can compute the total masses of the two isotopes in the atmosphere at time $t$:
\begin{equation}                                   \label{eq:Cisos}
  \Carbon{13}(t) = \frac{R(t)}{1 + R(t)}C(t), \quad
  \Carbon{12}(t) = \frac{1}{1+ R(t)}C(t).
\end{equation} 

We are now ready to analyze the data.  We employ four (related) methods which are described in detail below.

\subsection{Single time varying reservoir} 
\label{SEC:SingleBox}

\begin{figure}
\begin{framed}
\begin{center}
\includegraphics[width=.95\textwidth]{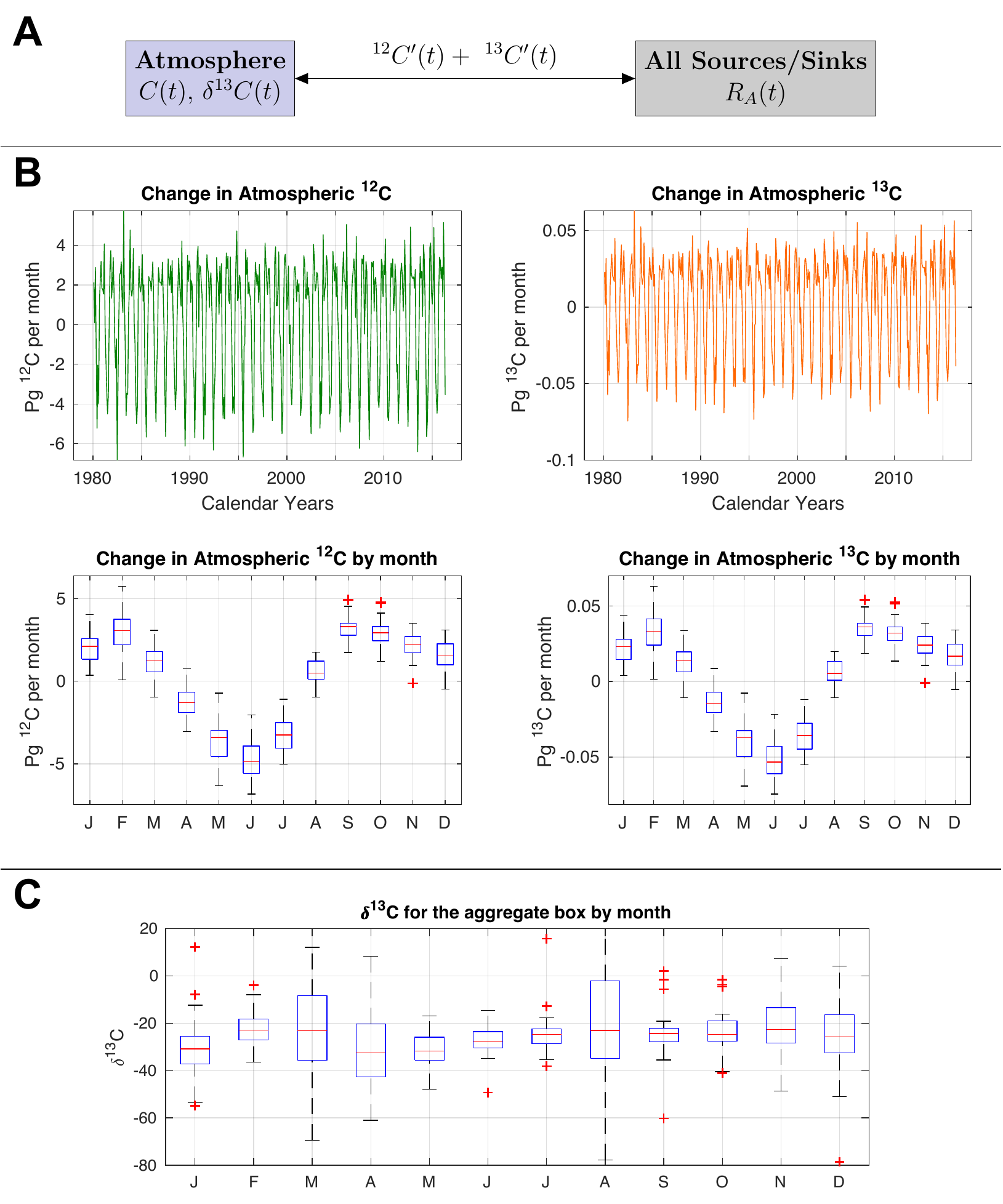}
\caption{Extremely simplified model in which all the carbon reservoirs are aggregated. \textbf{A:} Model schematic.  \textbf{B:} Time series for the monthly change in $\Carbon{12}$ and $\Carbon{13}$ and box and whisker plots for each month over the time series. On each box, the red line indicates the median, and the bottom and top edges of the box indicate the 25th and 75th percentiles, respectively. The whiskers extend to the most extreme data points not considered outliers, and the outliers are plotted individually using the `+' symbol. \textbf{C:} The $\delC$ ratio for the aggregate box depicted with a box and whisker plot for each month over the time series. }
\label{fig:Single-Reservoir-MLO}
\end{center}
\end{framed}
\end{figure}

As a first pass we suspend our knowledge of the physical constraints of the carbon cycle and consider a single carbon reservoir whose isotope ratio varies with time to produce the isotope ratio observed in the atmosphere.  This concept is illustrated in Figure~\ref{fig:Single-Reservoir-MLO}A.  

At each time step, we need to know the amount of carbon entering or leaving the atmosphere at at what isotopic ratio. The instantaneous rates at which carbon-13 and carbon-12 are flowing into the atmosphere are given by 
$\Carbon{13}^\prime(t)$ and $\Carbon{12}^\prime(t)$, respectively.  We approximate these rates by taking the difference between consecutive months, i.e.
\[\Carbon{12}'(t)\approx\frac{\Carbon{12}(t)- \ \Carbon{12}(t-1)}{t-(t-1)},\quad \Carbon{13}'(t)\approx\frac{\Carbon{13}(t)- \ \Carbon{13}(t-1)}{t-(t-1)}.\]
These approximate rates are graphed in Figure~\ref{fig:Single-Reservoir-MLO}B.  The time series are highly oscillatory, so we also present the data in a box and whisker plot by month.  In this format, the seasonal cycle in both $\Carbon{12}$ and $\Carbon{13}$ is apparent with a draw down in atmospheric carbon in the months of April through July and net emission during the remaining months. It is interesting to note that between September and February, the rate of increase (on average) slows down, perhaps due to the growth of the southern hemisphere biosphere.

The instantaneous ratio $R_A(t)$ of carbon-13 to carbon-12 in our conceptual aggregate reservoir is given by
\begin{equation*}
  R_A(t) = \frac{\Carbon{13}^\prime(t)}{\Carbon{12}^\prime(t)},
\end{equation*}
which can be expressed in the standard per mil notation as 
$\delta_A(t)$ given by inverting equation~\eqref{eq:R} to obtain
\begin{equation*}
  \delta_A(t) = 1000(R_A(t)/R_R - 1).
\end{equation*}
We emphasize that $\delta_A(t)$ is the carbon-13 to carbon-12 ratio (expressed in the ``per mil'' notation of formula~\eqref{eq:delC}) at time $t$ for the single reservoir supplying carbon to the atmosphere illustrated in Figure~\ref{fig:Single-Reservoir-MLO}A.  This ratio is shown in Figure~\ref{fig:Single-Reservoir-MLO}C. 

The ratio $\delta_A(t)$ is highly sensitive to the times that $\Carbon{12}'(t)$ and $\Carbon{13}'(t)$ cross from positive to negative or vice versa, resulting in nonphysical blowup of the signature if these crossings are not synchronous.  For this reason, we do not present the time series for $\delta_A(t)$ in Figure~\ref{fig:Single-Reservoir-MLO}C, and instead only present the box and whisker plot by month. This blowup feature is, of course, a major flaw in the assumption that the atmospheric carbon can be explained by one carbon reservoir. However, the medians exhibit a slight seasonal cycle, with times where the atmospheric carbon is decreasing having slightly more negative $\delC$ signatures than at times where the atmospheric carbon is increasing. This supports the argument that the observed draw down in atmsopheric carbon in the Mauna Loa data set is due to the growth of plants in the northern hemisphere.

\subsection{Biotic and abiotic reservoirs} \label{sect:MLO}

\begin{figure}
\begin{framed}
\begin{center}
\includegraphics[width=.95\textwidth]{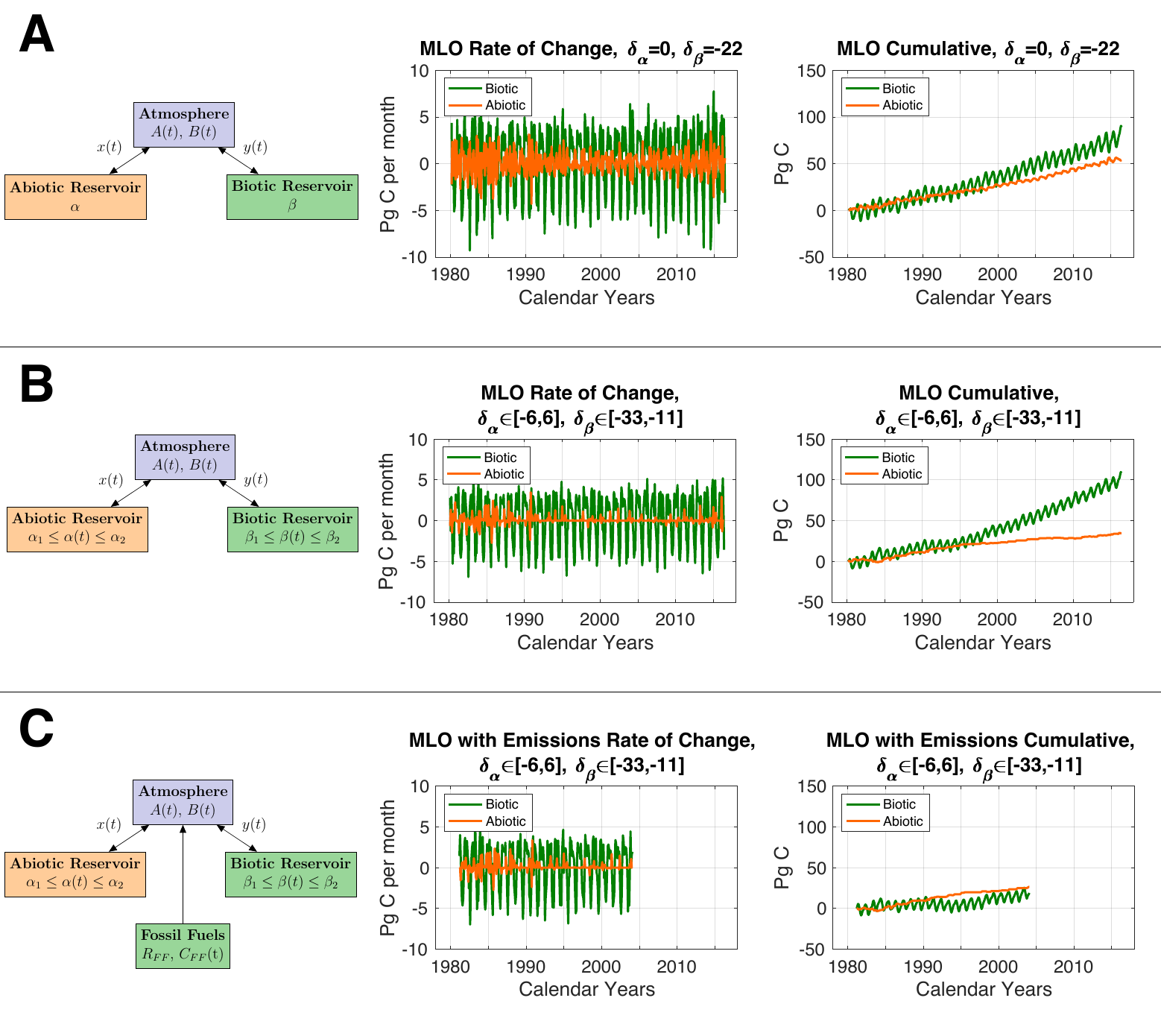}
\caption{Analysis with two carbon reservoirs having different $\delC$ levels. In each row, the left image is the model schematic, the middle image is the instantaneous rates from each reservoir (where positive indicates emission to the atmosphere) and the right image is the cumulative contribution from each reservoir. \textbf{Row A:} gives the results for fixed isotope ratios in the reservoirs (See Section~\ref{sect:MLO-fixed} for more details), \textbf{Row B:} gives the results for bounded, variable isotope ratios in the reservoirs (See Section~\ref{sect:MLO-range} for more details), and \textbf{Row C:} is the same as Row B but specifically accounts for fossil fuels emissions (See Section~\ref{sect:MLO-FF} for more details). }
\label{fig:IMG-biotic-abiotic}
\end{center}
\end{framed}
\end{figure}

Because of the singularities in $\delC$ signature as atmospheric carbon switches from increasing to decreasing (or vice versa), we see that a single aggregate box is insufficient to model the exchange of carbon between terrestrial reservoirs and the atmosphere.  The next obvious attempt should then use two boxes.  As described above, photosynthesis incorporates carbon-12 more easily than carbon-13 into the organic compounds produced by the process.  Therefore, the $\delC$ value of the plant material will be lower than that of the surrounding atmosphere.  This lower value will be passed on up the food chain to herbivores and predators and eventually to fossil fuels, which are the remains of biological materials long buried.

A reservoir of carbon that traces its origin back to photosynthesis, such as a vein of coal or the leaf liter on a forest floor, we call a \textit{biotic} reservoir.  A reservoir that has not originated through photosynthesis, such as the gases emitted by a volcano, we call an \textit{abiotic} reservoir.  These two types of sources and sinks are distinguished by differing $\delC$ levels. Typical  $\delC$ values for biotic carbon range from $-11\permil$ and lower, depending on the type of photosynthesis in the original organic matter and differing decay processes \cite{Farquhar,Rounick1986}. Conversely, physical processes such as the formation of calcium carbonate or volcanic release tend to leave samples with a $\delta^{13}$C signature closer to the standard of $0 \permil$ (for example  \cite{Kohler, Sigman, Ciais1995a, Francey1995}).

\subsubsection{Constant isotope ratios} \label{sect:MLO-fixed}

To refine our analysis we replace the single reservoir having large, nonphysical variations in the 
$\delC$ ratio with two reservoirs, each with a fixed 
$\delC$, one biotic and the other abiotic, as illustrated in Figure~\ref{fig:IMG-biotic-abiotic}A.  The ratio of carbon-13 to total carbon in the abiotic reservoir is denoted by $\alpha$ while that ratio in the biotic reservoir is denoted by $\beta$.  

Note that we have now adopted yet a third metric to measure the carbon isotope content.  The variable $R$ described in the introduction is the ratio of carbon-13 to carbon-12 in a particular sample.  In the literature, that ratio is usually described by the variable
$\delC$ as given in equation~\eqref{eq:delC}.  We are now using instead of either of these metrics the more mathematically natural ratio of carbon-13 to total carbon.  For example, if the abiotic reservoir has a ratio $R_a$ of carbon-13 to carbon-12, then the ratio $\alpha$ of carbon-13 to total carbon is given by
\begin{equation*}
  \alpha = \frac{R_a}{1 + R_a} .
\end{equation*}

To simplify the notation, we introduce new variables for the time rate of change of the total masses of the carbon-13 and carbon-12 in the atmosphere:
\begin{equation*}
  A(t) = {\Carbon{13}}^\prime(t), \quad
  B(t) = {\Carbon{12}}^\prime(t),
\end{equation*} 
where $\Carbon{13}(t)$ and $\Carbon{12}(t)$ are defined in equations~\eqref{eq:Cisos}.

We now let $x(t)$ be the rate at which carbon flows into the atmosphere from the abiotic reservoir and let $y(t)$ be the rate at which carbon flows into the atmosphere from the biotic reservoir.  These variables are related as follows:
\begin{equation*}
\begin{aligned}
  \alpha x + \beta y &= A\\
               x + y &= A + B
\end{aligned}
\end{equation*}
These equations simply state that the amounts of carbon-13 from each reservoir 
($\alpha x$ and $\beta y$) must sum to the total change in carbon-13 and that the amounts of total carbon from each pool 
($x$ and $y$) must sum to the total change in carbon ($A+B$). 

The goal is to use the data to determine the rates 
$x(t)$ and $y(t)$ at which the carbon is flowing in and out of the reservoirs.  The data give us 
$A(t) = {\Carbon{13}}^\prime(t)$ and 
$B(t) = {\Carbon{12}}^\prime(t)$ as described in Section 2\ref{SEC:SingleBox}.

As long as $\alpha\not=\beta$ we can invert this linear system to get the amounts of carbon from each pool in terms of the data and the concentrations $\alpha$ and $\beta$:
\begin{equation}
\begin{aligned}
x(t)&=\frac{A(t)-\beta(A(t)+B(t))}{\alpha-\beta}\\
y(t)&=\frac{\alpha(A(t)+B(t))-A(t)}{\alpha-\beta}.
\label{eq:x-y-fixed}
\end{aligned}
\end{equation}

The plots of $x(t)$ and $y(t)$ are shown in  the middle figure of Row A in Figure~\ref{fig:IMG-biotic-abiotic}.  In this figure, we take $\alpha$ and $\beta$ ratios so that the $\delC$ of the respective reservoirs are $\delta_\alpha=0\permil$ and $\delta_\beta=-22\permil$.  Like the $\Carbon{12}'(t)$ and $\Carbon{13}'(t)$ plots from Figure~\ref{fig:Single-Reservoir-MLO}B, these time series are highly oscillatory.  For this reason, we've also included the cumulative contributions in the right figure in Row A of Figure~\ref{fig:IMG-biotic-abiotic}.  

In the figure depicting the cumulative contributions from the biotic and abiotic pools, we can clearly see a seasonal cycle in both reservoirs.  Although a seasonal cycle is expected from the biotic reservoir, the behavior of the abiotic reservoir cannot be readily biologically explained.  We suspect that the abiotic cycle is present due to the underlying oscillatory behavior of $A(t)$ and $B(t)$ and the fact that $\alpha$ and $\beta$ are fixed.

Positive values in this plot indicate net emission to the atmosphere. We see that over the course of the times series both reservoirs were net emitters under this framework of analysis.  Over the the roughly 35 year period there was a net emission of about 80 Pg C from biotic pools and 50 Pg C from abiotic pools.

We have conducted this analysis with nine biologically and physically motivated different combinations for $\alpha$ and $\beta$.  The results of these different runs of the algorithm are presented the appendix, but the results described above are robust to biologically  and physically relevant ratios for the reservoirs.

\subsubsection{Variable isotope ratios} \label{sect:MLO-range}

It is unlikely that whatever reservoirs are active at various times will always have the same isotope ratios.  We would like the data to determine the ratios in the two boxes shown in Figure~\ref{fig:IMG-biotic-abiotic} rather than fixing the $\delC$ \emph{a priori}.
  
As we did in Section~2\ref{SEC:SingleBox}, we allow the isotope ratios to vary in the reservoirs, but now we restrict those levels between bounds.  The ratio of carbon-13 to total carbon in the abiotic reservoir is allowed to vary only between 
$\alpha_1$ and $\alpha_2$, while that ratio in the biotic reservoir is allowed to vary only between 
$\beta_1$ and $\beta_2$.  Denoting by 
$\alpha(t)$ the carbon-13 to total carbon ratio in the abiotic box and by 
$\beta(t)$ that ratio in the biotic reservoir, we have
\begin{equation*}
  \beta_1 \le \beta(t) \le \beta_2  <
  \alpha_1 \le \alpha(t) \le \alpha_2 .
\end{equation*}
Call the rectangular region denoted by these bounds $R$.

The contributions from the abiotic and biotic reservoirs are calculated the same way as in the fixed ratio analysis and $x(t)$ and $y(t)$ are given as in Equation~\eqref{eq:x-y-fixed}, now depending on unknowns $\alpha$ and $\beta$. We employ an occam's-razor approach to determining $\alpha$ and $\beta$ at each time step.  In absence of other information, this approach is desirable because it returns the smallest amount of carbon  from each reservoir needed to create what has been observed in the atmosphere. 
It can be shown that the following minimization process guarantees the direction of the flow---regardless of actual magnitudes exchanged, sources remain sources and sinks remain sinks.

\begin{figure}
\begin{framed}
\begin{center}
\includegraphics[width=.42\textwidth]{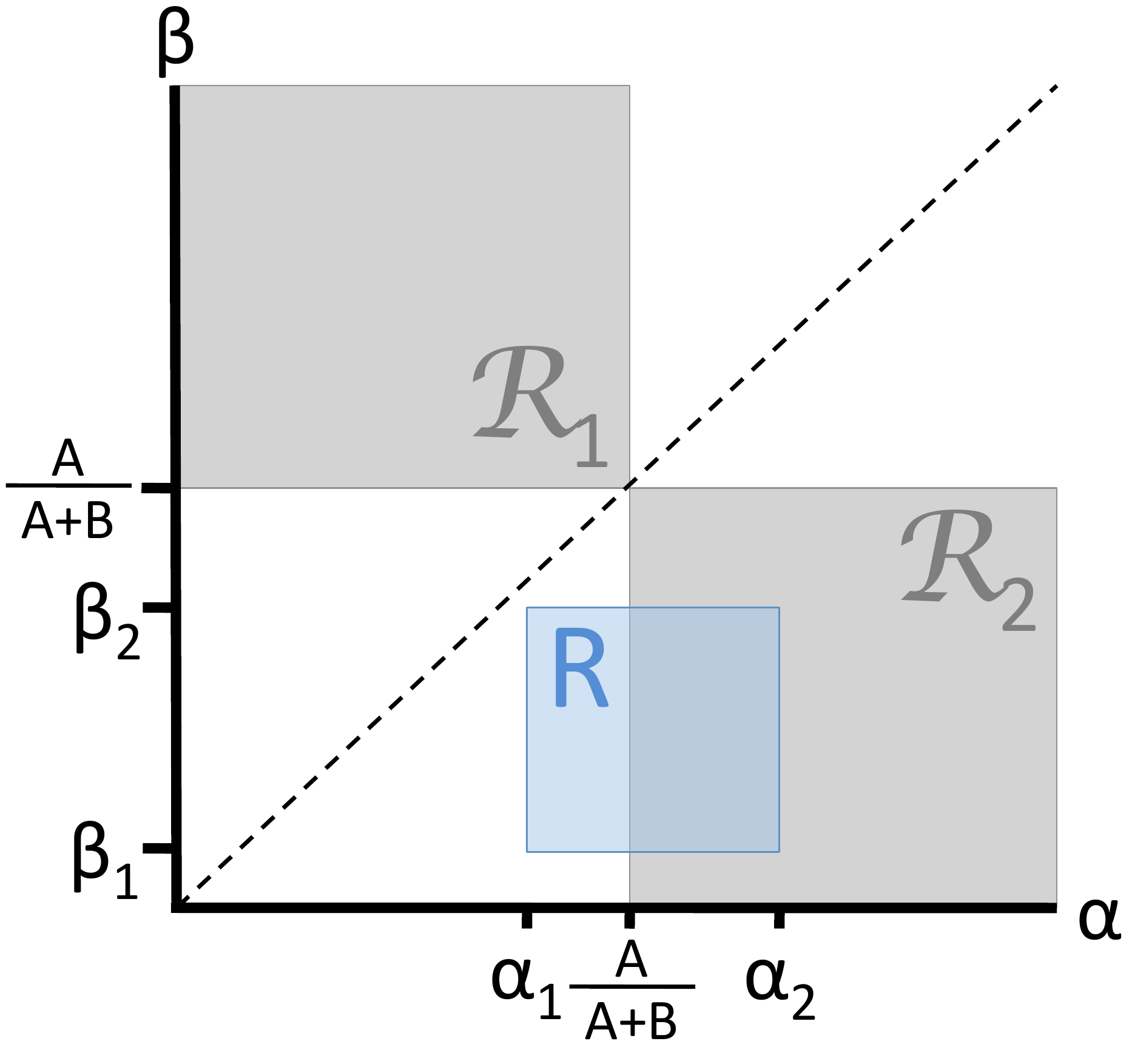} \hspace{1cm}
\includegraphics[width=.42\textwidth]{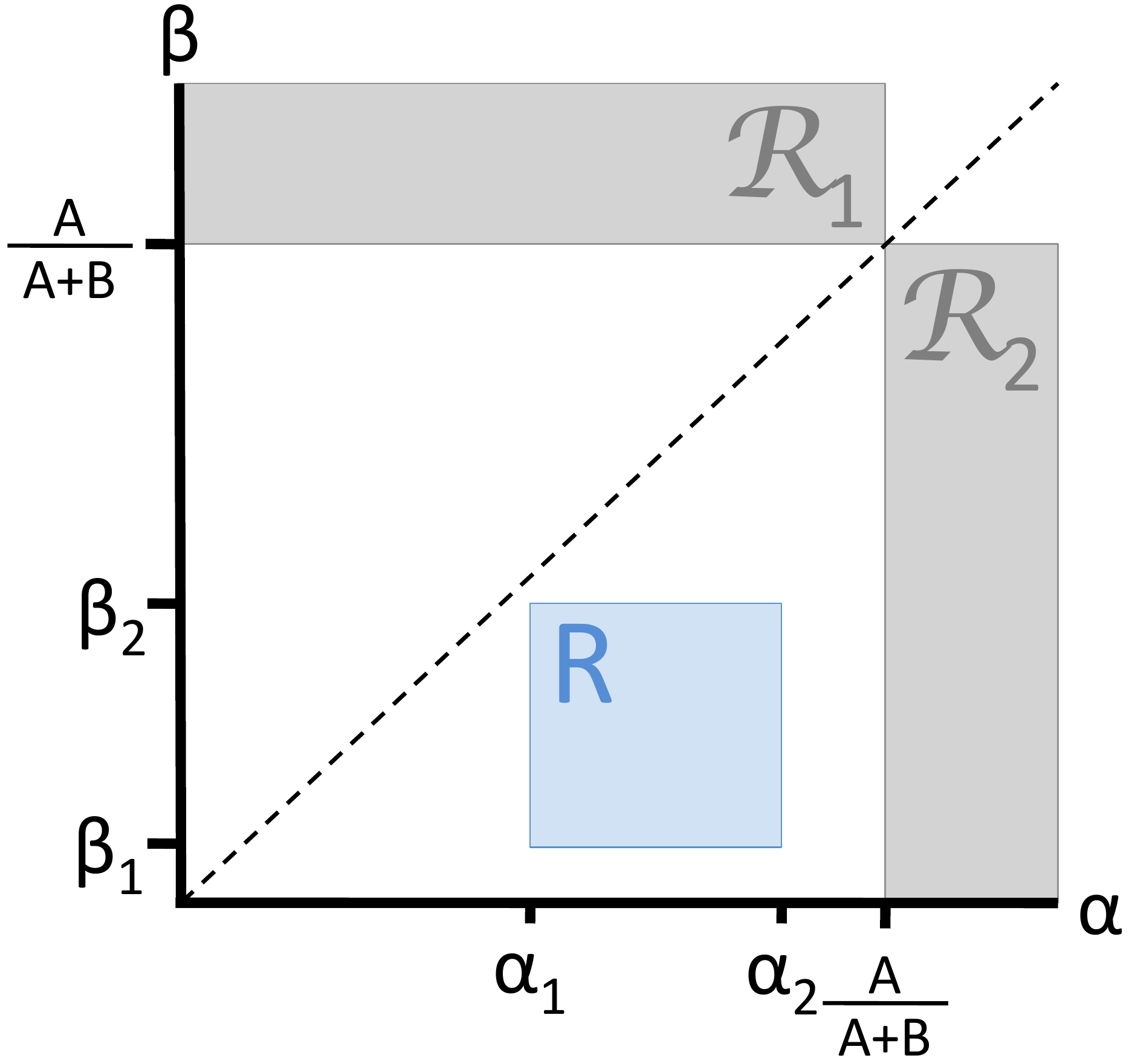}
\caption{The region of minimization, $R$, and the region on which the cost function is constant, $\mathcal R=\mathcal R_1\cup\mathcal R_2$, for two different scenarios of atmospheric data.   When $R\cap\mathcal R_2$ is nonempty, the minimum of the cost function is not unique, motivating further conditions for determining $\alpha$ and $\beta$. The dashed line indicates $\beta=\alpha$. }
\label{fig:regions}
\end{center}
\end{framed}
\end{figure}

The cost function that minimizes the total amount of carbon for each time step is the sum of the absolute values of carbon from each pool
\[F(\alpha,\beta)=\left|x(\alpha,\beta)\right|+\left|y(\alpha,\beta)\right|.\]
Because this cost function $F$ is constant in the region
\[\mathcal R=\underbrace{\left\{(\alpha,\beta):\alpha\leq\frac{A}{A+B},\ \beta\geq\frac{A}{A+B}\right\}}_{\mathcal R_1} \cup \underbrace{\left\{(\alpha,\beta):\alpha\geq\frac{A}{A+B},\ \beta\leq\frac{A}{A+B}\right\}}_{\mathcal R_2},\]
 the minimum of $F$ in the region $R$ is not unique if $R\cap\mathcal R_2$ is nonempty. For clarity, a depiction of these regions is given in Figure~\ref{fig:regions}.
 
To address the issue of a nonunique minimum of $F$, we give the following conditions for determining $\alpha$ and $\beta$ based on the instantaneous atmospheric data ($A/(A~+~B)$):
\begin{enumerate}
\item If $\frac{A}{A+B}<\beta_1$ or $\frac{A}{A+B}>\alpha_1$, then $F$ has a unique minimum at $\beta = \beta_1$ and $\alpha=\alpha_2$, giving $x=x(\alpha_2,\beta_1)$ and $y=y(\alpha_2,\beta_1)$.
\item  \textbf{Motivation:} If the instantaneous carbon ratio falls between the ratios for the abiotic and biotic reservoirs, then both pools must have contributed to atmospheric carbon in that time step. Take the highest possible concentration from the biotic pool and the lowest possible concentration from the abiotic pool.
	\begin{enumerate}
	\item If $\beta_2<\frac{A}{A+B}<\alpha_1$, let $\beta = \beta_2$ and $\alpha=\alpha_1$. Then $x=x(\alpha_1,\beta_2)$ and $y=y(\alpha_1,\beta_2)$.
	\end{enumerate}
\item \textbf{Motivation:} If the instantaneous carbon ratio corresponds to a possible ratio of one of the two carbon pools, then the simplest explanation which minimizes the flow is given by attributing all of the atmospheric exchange to that pool and none to the other.
	\begin{enumerate} 
	\item If $\alpha_1\leq\frac{A}{A+B}\leq\alpha_2$,  let $\alpha = \frac{A}{A+B}$. Then $x=A+B$ and $y=0$ regardless of the value of $\beta$.  
	\item If $\beta_1\leq\frac{A}{A+B}\leq\beta_2$,  let $\beta = \frac{A}{A+B}$. Then $y=A+B$ and $x=0$ regardless of the value of $\alpha$.  
	\item To determine $\beta$ for the case in (a) (or $\alpha$ for the case in (b)), we choose $\beta$ (or $\alpha$) so that $(\alpha,\beta)$ lies on the line that connects the antipodal corners of the rectangle $R$ $(\alpha_1,\beta_2)$ and $(\alpha_2,\beta_1)$, i.e. $\alpha$ and $\beta$ must satisfy
	\[\beta-\beta_2=\frac{\alpha_1-\alpha_2}{\beta_2-\beta_1}(\alpha-\alpha_1).\]
	Determining $\beta$ (or $\alpha$) in this way does not affect the value of $x$ or $y$ but does ensure that $\beta(A,B)$ (or $\alpha(A,B)$) is a continuous function of $A$ and $B$ for the minimization criteria.
	\end{enumerate}
\end{enumerate}
Pseudocode for this minimization algorithm is given in Appendix \ref{Sect:Pseudocode}.

The result of implementing this minimization algorithm over the time series is depicted in Row B of Figure~\ref{fig:IMG-biotic-abiotic}. In these figures we take the bounds on $\alpha$ and $\beta$ so that $\delta_\alpha\in[-6\permil,6\permil]$ and $\delta_\beta\in[-33\permil,-11\permil]$.  The range for the biotic box covers the range of found in plants with either C3 or C4 photosynthesis \cite{Farquhar}.  Again, these rate plots are highly oscillatory; however, we do see that the change in the abiotic reservoir is greatly diminished compared to the results with fixed isotope ratios.

In the rightmost plot of Row B in Figure~\ref{fig:IMG-biotic-abiotic}, we show the cumulative contributions from the biotic and abiotic reservoirs.  In contrast with the fixed ratios case, we see a seasonal cycle only from the biotic reservoir. The loss of the seasonal cycle in the abiotic reservoir gives us confidence that this method is not including extraneous information.

Again, positive values indicate net emission to the atmosphere.  Over the period analyzed, results indicate a net emission from the biotic box of about 100 Pg C and a net emission from the abiotic box of about 30 Pg C. The large amount of contribution from the abiotic reservoir is striking, since, presumably, this increase in atmospheric carbon is due overwhelmingly to the burning of fossil fuels, a biotic source. Perhaps this is explained by the effect of mixing of carbon dioxide, between the atmosphere and the ocean with a net effect of biotic carbon being absorbed, but abiotic carbon being released.

We have conducted this analysis with nine different combinations for the bounds on $\alpha$ and $\beta$.  The results of these different runs of the algorithm are presented in the appendix, but the results described above are robust to biologically and physically relevant bounds on the ratios for the reservoirs.

\subsubsection{Adding a net carbon source} \label{sect:MLO-FF}

We'd like to further constrain the behavior in the biotic and abiotic reservoirs by including information about known sources (or sinks) of carbon.  This information is readily incorporated into the analysis provided that the total carbon emitted (or sequestered) and its $\delC$ ratio are known to a reasonable degree of accuracy.

To illustrate incorporating a carbon source into the analysis, we will use data from the Carbon Dioxide Information Analysis Center \cite{Blasing} on United States carbon emissions from 1981 to 2003.  Readers interested in how to incorporate a sink should refer to Section~3c(\ref{sect:peatlands}).

\begin{figure}
\begin{framed}
\begin{center}
\includegraphics[width=.9\textwidth]{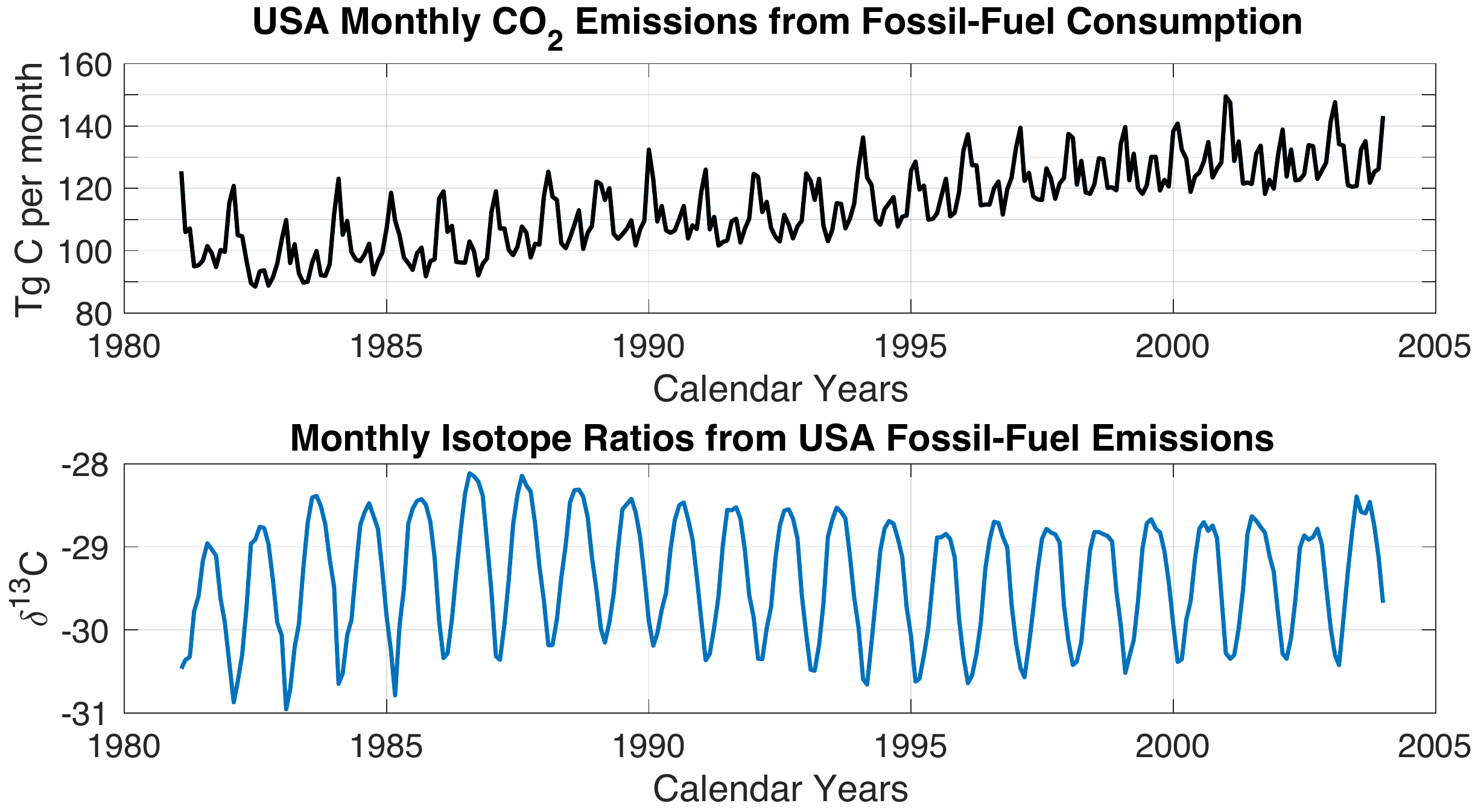}
\caption{USA carbon emission data from the Carbon Dioxide Information Analysis Center \cite{Blasing}.}
\label{fig:CDIAC}
\end{center}
\end{framed}
\end{figure}

We selected the data in \cite{Blasing} for our analysis because it provides monthly carbon emissions due to fossil fuel consumption in the United States in teragrams carbon along with monthly $\delC$ estimates.  The data are shown in Figure~\ref{fig:CDIAC}. If you must work with reservoir data which has different temporal spacing than the original atmospheric data, you could interpolate between points with an appropriate spline function.  Unfortunately, this data ends in December 2003. The Carbon Dioxide Information Analysis Center do provide annual average emissions to more current dates, but incorporating the annual average emissions data into the seasonal Mauna Loa data may affect the seasonal cycle in unexpected in spurious ways.  In the annual average data were the only data available, it would be better to use annual average data from Mauna Loa instead of the monthly averages.

To prepare the data for analysis, we convert the total carbon to petagrams C by dividing by 1000.  Because the data is presented as monthly rates, we will be able to transform the data directly into $\Carbon{12}'(t)$ and $\Carbon{13}'(t)$ without first finding $\Carbon{12}(t)$ and $\Carbon{13}(t)$. Let $C_{FF}'(t)$ denote the monthly fossil fuel emission data in petagrams carbon and denote CDIAC $\delC$ data as
\[\delta_{FF}(t)=1000\times\left(\frac{\frac{\Carbon{13}'_{FF}(t)}{\Carbon{12}'_{FF}(t)}}{R_R}-1\right)\]
Then the isotope ratio is given by
\begin{equation*}                                                                            
  R_{FF}(t) = R_R\cdot(1 + 0.001\cdot\delta_{FF}(t)),
\end{equation*}
Now that we have the ratio $R_{FF}(t)$ of $\Carbon{13}'(t)$ and $\Carbon{12}'(t)$ and the total emissions $C'_{FF}(t)$, we can compute the total emissions of the two isotopes time $t$:
\begin{equation*}                                 
  \Carbon{13}'_{FF}(t) = \frac{R_{FF}(t)}{1 + R_{FF}(t)}C'_{FF}(t), \quad
  \Carbon{12}'_{FF}(t) = \frac{1}{1+ R_{FF}(t)}C'_{FF}(t).
\end{equation*} 
To simplify the notation, we introduce new variables :
\begin{equation*}
  A_{FF}(t) = {\Carbon{13}'_{FF}}(t), \quad
  B_{FF}(t) = {\Carbon{12}'_{FF}}(t).
\end{equation*} 

As in the previous two sections, let $x(t)$ be the rate at which carbon flows into the atmosphere from the abiotic reservoir and let $y(t)$ be the rate at which carbon flows into the atmosphere from the biotic reservoir.  These variables, along with the fossil fuel data, are related as follows:
\begin{equation*}
\begin{aligned}
  \alpha x + \beta y +A_{FF} &= A\\
               x + y + A_{FF} + B_{FF} &= A + B
\end{aligned}
\end{equation*}
These equations simply state that the amounts of carbon-13 from each reservoir and the burning of fossil fuels
($\alpha x$, $\beta y$, and $A_{FF}$ ) must sum to the total change in carbon-13 and that the amounts of total carbon from each pool and the total amount emitted from burning fossil fuels
($x$, $y$, and $A_{FF}+B_{FF}$) must sum to the total change in carbon ($A+B$). Moving all of the knowns to the righthand side of the equal sign gives
\begin{equation*}
\begin{aligned}
  \alpha x + \beta y &= \tilde A\\
               x + y &= \tilde A + \tilde B
\end{aligned}
\end{equation*}
where
\begin{equation*}
  \tilde A = A-A_{FF} , \quad
  \tilde B= B- B_{FF}.
\end{equation*} 
As before, we invert the system to find $x$ and $y$ as functions of the data ($\tilde A$ and $\tilde B$) and parameters ($\alpha$ and $\beta$).

The results of this analysis are presented in Row C in Figure~\ref{fig:IMG-biotic-abiotic}. Note that we present results only for the dates on which the data sets coincide, i.e. January 1981 to December 2003.  We use the same bounds for the biotic and abiotic isotope ratios as we did in Section~2c(\ref{sect:MLO-range}).  

In the rightmost plot in Row C, we see that accounting for the USA fossil fuel emissions results in a decrease in the cumulative contribution from the biotic reservoir by about 35 Pg carbon (relative to the same time in the biotic times series in Row B).  This is expected because the isotopic ratio of fossil fuels has a biotic signature.  

Even though the fossil fuel data has a seasonal cycle, it has not canceled out the  seasonal cycle in $y$.  This is because the amplitude for the oscillations in the USA fossil fuel emissions are about two orders of magnitude smaller than the amplitude for the oscillations in the Mauna Loa data.

Comparing the rightmost plot of Row C with that of Row B, we see that the incorporation of the emissions data with the Mauna Loa data has significantly decreased the contribution from the biotic reservoir, but the contribution from the abiotic reservoir is relatively unchanged.  The decrease in the biotic source is expected because we have accounted for some of those emissions with the fossil fuel data from CDIAC.  The persistent source of abiotic carbon is more mysterious, as we are unaware of any known abiotic source of this magnitude over this time period.  Possible explanations include volcanic emissions or that the continuous interchange of carbon between the ocean and the atmosphere creates the appearance of a net abiotic source, although this is far from an exhaustive list. More analytical and empirical work is need to confirm or rule out these hypotheses.

We have conducted this analysis with nine different combinations for the bounds on $\alpha$ and $\beta$.  The results of these different runs of the algorithm are presented in the appendix, but the results described above are robust to biologically and physically relevant bounds on the ratios for the reservoirs.

\section{Paleoclimate Carbon Data} \label{sect:paleo-sect}

The frameworks described above can be used to study carbon sources and sinks during any time period as long as a suitable time series of the carbon data can be obtained.  To demonstrate the robustness of these methods we additionally investigate paleocarbon data provided by Schmitt \emph{et al} \cite{Schmitt}  and Monnin~\emph{et al} \cite{Monnin} for the time period 20,000 years before present (20 kyr BP) to present.

The paleocarbon data comes from Antarctic ice cores \cite{Schmitt,Monnin}.  As snow is deposited on the Antarctic ice sheet and as the snow compacts to ice, tiny bubbles are trapped in the ice.  These bubbles provide a time series of samples of the past atmospheres.  The samples can be analyzed to determine both the atmospheric concentration of carbon dioxide and the relative abundance of carbon isotopes.

The data are derived from an analysis of the past atmospheric samples from Antarctic ice cores covering the last 20,000 years, during which time the ice sheets covering large parts of the Northern Hemisphere melted down to their current location, mostly in Greenland.  The concentration of atmospheric carbon increased during this period, while the isotopic composition varied as shown in the bottom plot of Figure~\ref{fig:data}.

\subsection{Transforming the data}
As with the Scripps data, the paleo carbon data are reported as a time series of concentrations of atmospheric carbon dioxide, given in parts per million by volume (ppmv), and $\delC$, as described above.  Unfortunately, due to challenges with collecting the data and the manner in which the air bubbles form, the times series for atmospheric carbon dioxide concentration is not equally spaced.  Furthermore, time between samples may span several hundred years which is not ideal for any of the models described in the previous section. The atmospheric concentration data are provided by Monnin \cite{Monnin} and the isotopes ratios are provided by Schmitt \emph{et al} \cite{Schmitt}.

To interpolate between the data points in the atmospheric concentration \cite{Monnin}, we use {\sc Matlab}'s smoothing cubic spline function \cite{Reinsch} \texttt{csaps} with smoothing parameter $p=10^{-8}$.  This {\sc Matlab} function returns a cubic spline $f$ which minimizes the objective function
\[p\underbrace{\sum_{j=1}^{n}|y_j-f(x_j)|^2}_{\text{error measure}} \ + \ (1-p)\underbrace{\int|D^2f(t)|^2dt}_{\text{roughness measure}}\]
where $n$ is the number of entries of $x$ and the integral is over the smallest interval containing all of the entries of $x$. We choose to use a smoothing cubic spline over the traditional cubic spline because the differences in neighboring data points are larger than one standard deviation of the data that went in to creating the point (points in the data set are the mean value of 4-6 neighboring samples) about 75\% of the time. Future work could include exploring how the interpolating function affects the results of the analysis, but that is not explored here.

Schmitt \emph{et al} construct an interpolation of the data for $\delC$ for every year in the 20,000 year time series \cite{Schmitt}.  However, their $\delC$ data is given as piecewise constant values because their data is given only to four significant figures.  We use their $\delC$ data in this analysis after fitting a smoothing cubic spline with {\sc Matlab}'s \texttt{csaps} function and smoothing parameter $p=10^{-8}$. 

For both data sets, we interpolate at each year in in the time series, resulting in vectors with 20,000 terms. From there, we calculate the $\Carbon{12}$ and $\Carbon{13}$ in the same manner that we did in Section~2\ref{sect:transforming}.  For $\Carbon{12}'(t)$ and $\Carbon{13}'(t)$ we have the added benefit that can actually compute the derivative because we are using a cubic spline.  We are now ready to analyze the data.  We will use the same four methods that we used on the Mauna Loa data.  The results are described in detail below.

\subsection{Single time varying reservoir} 

\begin{figure}
\begin{framed}
\begin{center}
\includegraphics[width=.95\textwidth]{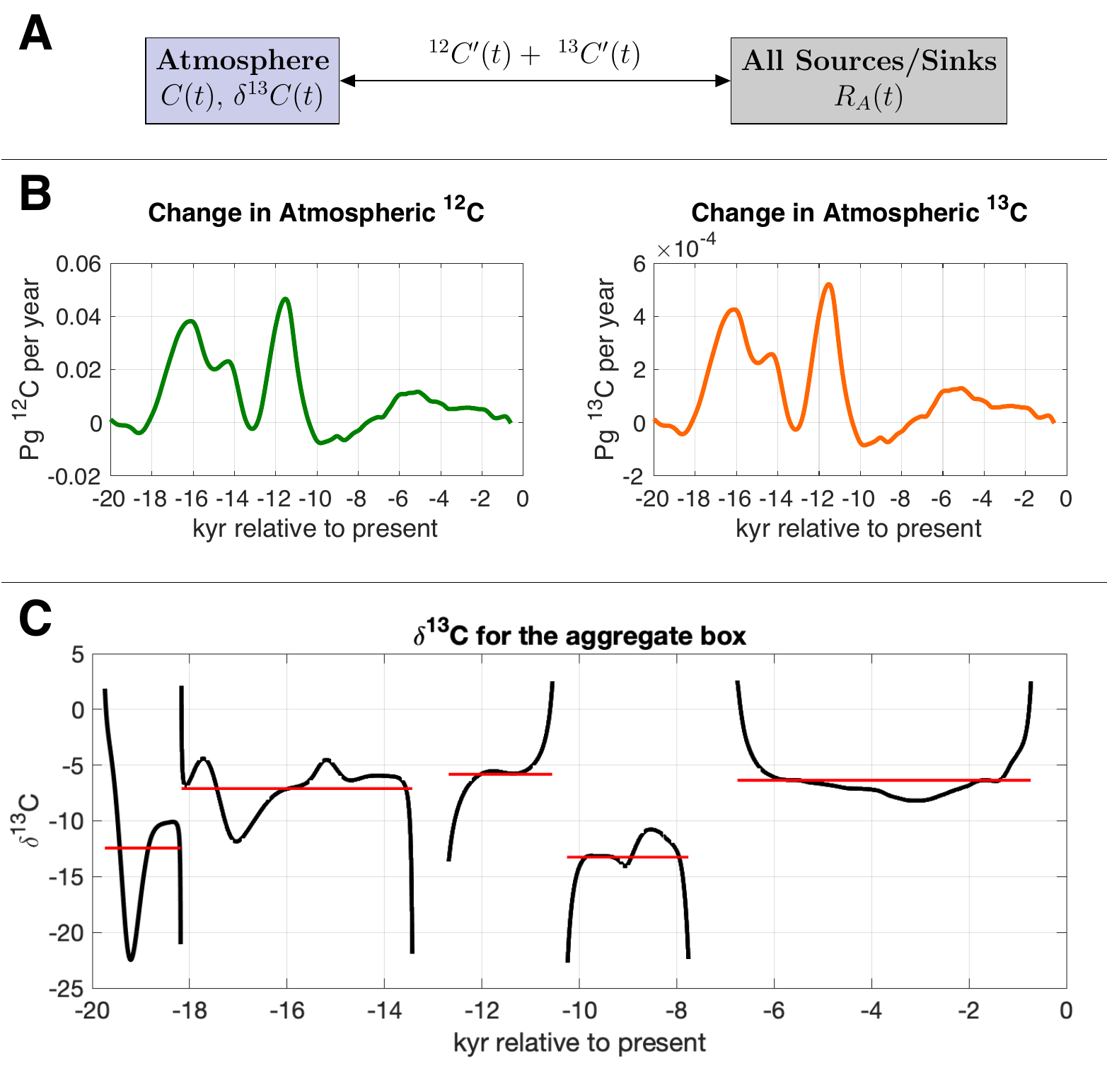}
\caption{Extremely simplified model in which all the carbon reservoirs are aggregated. \textbf{A:} Model schematic.  \textbf{B:} Time series for the annual change in $\Carbon{12}$ and $\Carbon{13}$. \textbf{C:} The $\delC$ ratio for the aggregate box with the parts of the time series that blow up removed. The horizontal red lines show the average $\delC$ value for that time interval. }
\label{fig:Single-Reservoir-Paleo}
\end{center}
\end{framed}
\end{figure}

When we consider the Schmitt and Monnin data in the model with a single time varying reservoir, we see that $\delta_A(t)$ becomes undefined at those times when the required concentration of carbon-12 in the aggregate box becomes zero, as seen in the spikes in Figure~\ref{fig:Single-Reservoir-Paleo}C.  This feature is, again, a major flaw in the assumption that the atmospheric carbon can be explained by one carbon reservoir.  However, an interesting observation emerges from this unrealistic assumption.  During times when the atmospheric concentration of carbon is increasing, the imaginary box, representing an aggregate reservoir combining all the carbon sources and sinks into a single source, has a 
$\delC$ value of approximately $-6\permil$.  However, between -20000 years and -18000 years before present and -10000 years and -8000 years before present, the periods when the atmospheric carbon was decreasing, the aggregate reservoir represents a single sink, with a 
$\delC$ value of about $-13\permil$.  The dynamics of the atmospheric carbon during these 2000 year periods appear to have been dominated by a biological sink, perhaps the growth of terrestrial vegetation as the ice sheets retreated..

\subsection{Biotic and abiotic reservoirs} \label{sect:Paleo}

\begin{figure}
\begin{framed}
\begin{center}
\includegraphics[width=.95\textwidth]{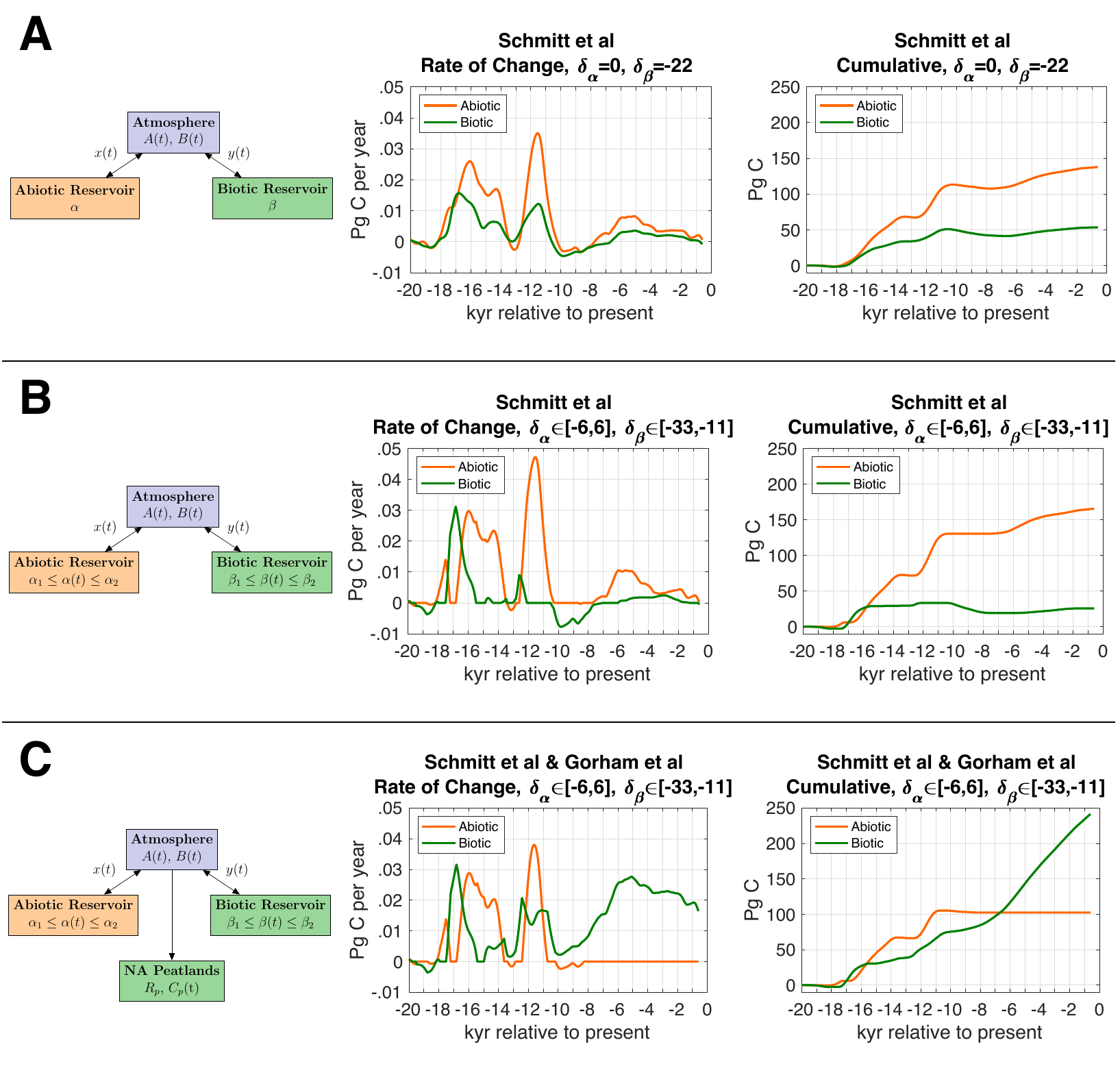}
\caption{Analysis with two carbon reservoirs having different $\delC$ levels. In each row, the left image is the model schematic, the middle image is the instantaneous rates from each reservoir (where positive indicates emission to the atmosphere) and the right image is the cumulative contribution from each reservoir. \textbf{Row A:} gives the results for fixed isotope ratios in the reservoirs (See Section~\ref{sect:Paleo-fixed} for discussion), \textbf{Row B:} gives the results for bounded, variable isotope ratios in the reservoirs (See Section~\ref{sect:Paleo-range} for discussion), and \textbf{Row C:} is the same as Row B but specifically accounts for growth of North American peatlands after glacial retreat (See Section~\ref{sect:peatlands} for discussion). }
\label{fig:IMG-biotic-abiotic-paleo}
\end{center}
\end{framed}
\end{figure}

As we did above with the Mauna Loa data, we now move from a single reservoir to two reservoirs, one biotic and one abiotic. As we describe below, we see general agreement with the conclusions drawn in Schmitt \emph{et al} that there was a net biotic sink during deglaciation, hypothesized to be related to the regrowth of the northern biosphere as the earth warmed and large areas formerly under ice were uncovered \cite{Schmitt}.

Remarkably, however, incorporating one known biotic sink---the regrowth of North American peatlands---more than accounts for this deficiency.  Apparently, from about 13,000 years ago to present, on balance carbon was flowing out of some unspecified biotic pools instead of into them. 

We have conducted this analysis with nine different combinations for the values or bounds on $\alpha$ and $\beta$.  The results are qualitatively robust and are presented in the appendix.

\subsubsection{Constant isotope ratios}  \label{sect:Paleo-fixed}

When we analyze the Schmitt \cite{Schmitt} and Monnin \cite{Monnin} data in the model with biotic and abiotic reservoirs and fixed isotope ratios, we see that the abiotic reservoir dominates over the whole time period.  This behavior is depicted in Row A of Figure~\ref{fig:IMG-biotic-abiotic-paleo}.  Under this framework of analysis, when there is a decrease in atmospheric carbon, we see that the biotic and abiotic boxes play comparable roles for duration and magnitude.  However, atmospheric increases are attributed to abiotic sources over biotic sources with, at times, a factor of three times as much, consistent with the widely accepted conclusion that the ocean was the major source of the increase in atmospheric carbon since the last deglaciation.

The cumulative contribution of the abiotic and biotic reservoirs are net emissions of approximately 140 Pg C and 50 Pg C, respectively.

We have conducted this analysis with nine biologically and physically motivated different combinations for $\alpha$ and $\beta$ and the results are in the appendix. The results described above are robust to biologically and physically relevant ratios for the reservoirs.

\subsubsection{Variable isotope ratios} \label{sect:Paleo-range}

It is even more unlikely than the recent data that, over the past 20,000 years, whatever reservoirs were active at various times would always have the same isotope ratios, making the variable isotope analysis even more applicable.

The differences between the variable isotope case and the fixed isotope case for these data sets are more nuanced than with the Mauna Loa data. We see in Row B of Figure~\ref{fig:IMG-biotic-abiotic-paleo} that abiotic sources are still dominating over most of the course of the time series, again consistent with the ocean as the major source of carbon. An interesting exception is the biotic source peaking at 0.03 Pg C per year at 17 kyr BP while the abiotic source has decreased to zero.

The most striking difference between the variable and fixed isotope ratio analyses is that with the variable isotope ratios there are long intervals of time when the change in atmospheric carbon is completely explained by one of the reservoirs. In particular, we see that decreases in atmospheric carbon from 20 kyr BP to 18 kyr BP and 11 kyr BP to 7 kyr BP can be exclusively attributed to a biotic sink. This behavior during the latter period was remarked above as consistent with a growth of terrestrial vegetation during the rapid deglaciation.

The cumulative result of these changes is shown in the rightmost graph in Row B. Comparing this graph to the analogous graph in Row A, we see that using variable isotope ratios in the reservoirs yields a result indicating a larger role for the abiotic sources than does the fixed isotope analysis.

\subsubsection{Adding a net carbon sink}  \label{sect:peatlands}

\begin{figure}
\begin{framed}
\begin{center}
\includegraphics[width=.85\textwidth]{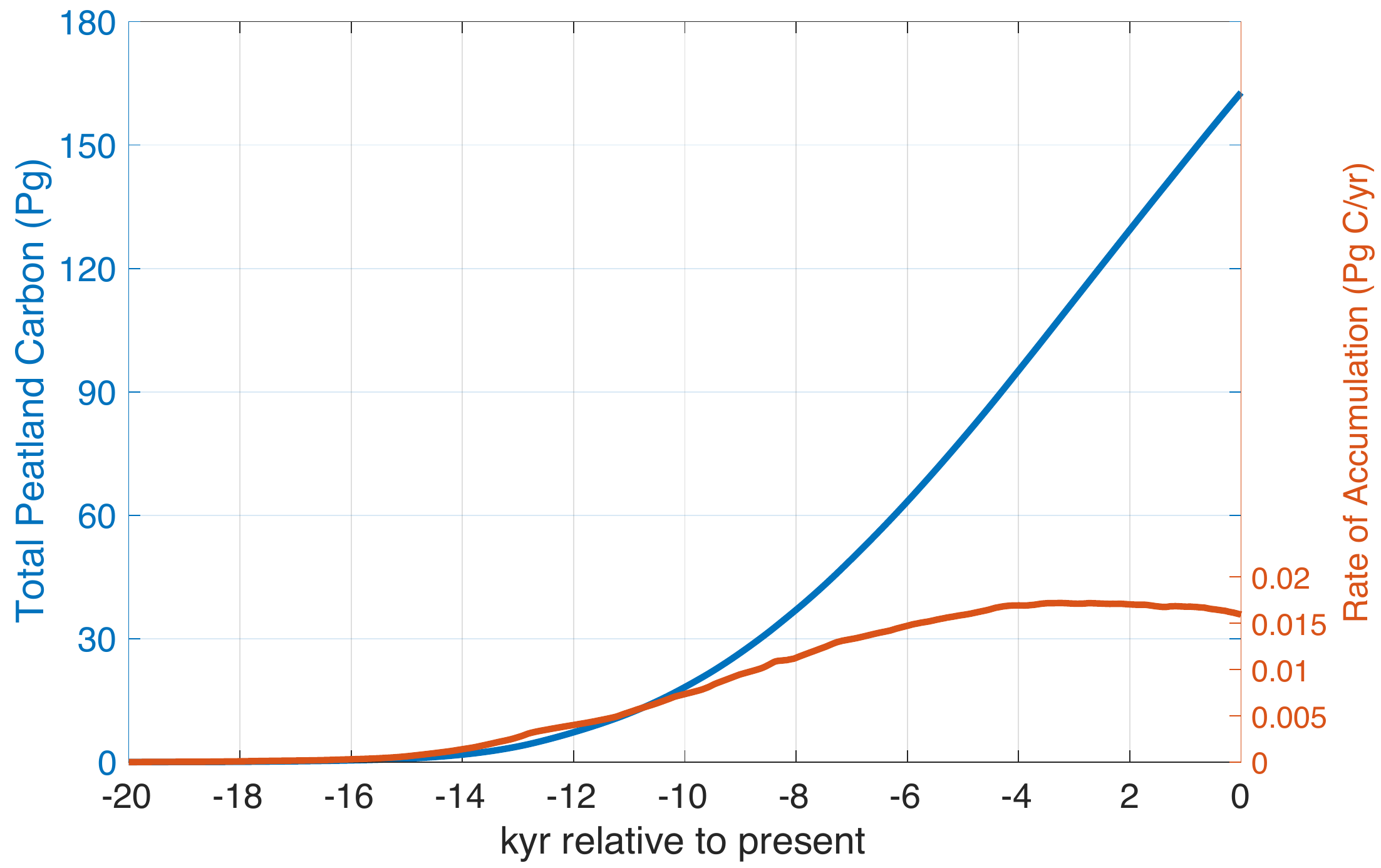}
\caption{North American peatland sequestration rate and cumulative carbon sequestered coomputed using theory from Gorham \emph{et al} \cite{Gorham12}.}
\label{fig:peatlands}
\end{center}
\end{framed}
\end{figure}

As before, we'd like to put further constraints on our unknown reservoirs; however this time we will use a carbon sink instead of a carbon source. A major carbon sink in the northern hemisphere during this time period was the growth of peatlands in land previously covered by glaciers \cite{Gorham07}.  The growth of North American peatlands after glacial retreat has been studied in depth and in this work we'll rely on several experimental and theoretical results of Eville Gorham and his collaborators \cite{Gorham91,Gorham07,  Gorham12}. 

In \cite{Gorham07,Gorham12}, the authors compile peatland locations, ages, and depths from over 2,000 North American peatland sites.  The authors then use this data to reconstruct the rate of carbon sequestration into North America peatlands taken as a whole.  To determine the amount of carbon sequestered into the peatlands over time and the rate at which it was sequestered, we rely on the theory developed in \cite{Gorham12} (see Table 1 in \cite{Gorham12} and discussion thereof).  The cumulative carbon stored in peatlands and rate of sequestration are depicted in Figure~\ref{fig:peatlands}. 

In this study we will use a constant $\delC$ ratio for the peatlands due to the lack of high resolution $\delta^{13}$C data for peatlands as a whole. Although studies of $\delC$ for individual peatlands have been conducted  (e.g. \cite{ Andersson, Aucour, Cristea,Kruger,  Skrypek,  Zhang,Zhong}), we propose that a constant $\delta^{13}$C value is adequate for a first-order understanding of carbon isotope composition of peatlands and sufficient for showing the benefit of the method. We use the mean $\delC$ from these studies, $-25\permil$.  Further studies may wish to investigate how this assumption on constant $\delC$ affects the results.

To analyze the data, we follow the framework laid out in Section~2c(\ref{sect:MLO-FF}). Again we are lucky because the reservoir data has the same temporal spacing as our atmospheric data and is presented as rates, allowing us to again find $\Carbon{12}'(t)$ and $\Carbon{13}'(t)$ without first finding $\Carbon{12}(t)$ and $\Carbon{13}(t)$.  Let
\begin{equation*}
  A_{p}(t) = {\Carbon{13}'_{p}}(t), \quad
  B_{p}(t) = {\Carbon{12}'_{p}}(t).
\end{equation*} 
denote the instantaneous net rates of carbon-13 and carbon-12 sequestered into North American peatlands.  Crucially, notice that because the flow of carbon is from the atmosphere to a sink, $A_p$ and $B_p$ are negative.  Then we will solve the system
\begin{equation*}
\begin{aligned}
  \alpha x + \beta y +A_{p} &= A\\
               x + y + A_{p} + B_{p} &= A + B
\end{aligned}
\end{equation*}
as in in Section~2c(\ref{sect:MLO-FF}) to find $x$ and $y$ after accounting for the peatland reservoir

The biotic and abiotic rates between and cumulative contribution to the atmosphere after accounting for peatland growth are plotted in Row C of Figure~\ref{fig:IMG-biotic-abiotic-paleo}.  In this figure we see that the peatland  sequestration rates are large enough to cause the slight biotic sink from Row B to become to a biotic source, making $y(t)$ positive for 13 kyr BP to present.  Qualitatively, this means that in order to account for the growth of North American peatlands, the biotic carbon reservoir must act as a large carbon source over this time period.  Conversely, after about 11 kyr BP, the abiotic contribution is negligible, meaning that one could explain all net changes in the atmosphere from 11 kyr BP to present by the growth of the peatlands and a yet undetermined biotic source. We note that this large biotic carbon source during the last 8000 years is consistent with the Ruddiman hypothesis that, during this period, human agriculture has been affecting the Earth's climate \cite{Ruddiman, Kaplan}.

\section{Discussion of the Methods} \label{sect:methods}

The first method proposed for studying changes in atmospheric carbon, where one attempts to explain the changes observed in the atmosphere with one aggregate carbon reservoir, can be useful provided one acknowledges its limitations.  While assuming one aggregate reservoir does cause blow-up in the isotope signature as atmospheric carbon changes from increasing to decreasing (or vice versa), we see that analysis of the $\delC$ signature for time intervals away from this blow-up may indicate predominant reservoirs acting in these intervals. This information could help scientists identify possible carbon sinks or sources, support current hypotheses, or motivate empirical work if the data suggest a previously unknown source or sink.  However, this method alone cannot provide a definitive account for atmospheric carbon exchange due to the uncertainties near blow-up.

The method of fixing the isotope ratios in the biotic and abiotic boxes and determining the flow is most similar to  other methods currently in the literature.  These methods partition the carbon budget into various flows between reservoirs and the atmosphere (at times with flows both in and out of the same reservoir) using mass-balance of carbon exchanged and fixed isotope ratios.  These methods  were initiated and championed by Berner in the early 1990's and 2000's \cite{Berner1991,Berner2001,Berner2003} and are still used in carbon budget analysis to this day \cite{Shields2017,Mills2019}. The main difference between the method proposed here and the ones currently in the literature is that here we consider only the net exchange with the atmosphere after each time interval instead of determining several flows both in and out of the atmosphere to different reservoirs. 

In any future analysis of the carbon budget (from any epoch), we propose using the third method which allows for variable isotope ratios for the biotic and abiotic reservoirs.  This method combines the mass-balance approach of the second method and the variable isotope ratios of the first method.  Furthermore, to use the method, we need not make any \emph{a priori} assumptions about the composition of the reservoirs except to set biologically and physically motivated bounds on their ratios.  The ease of adding known exchanges with the atmosphere, be they sinks or sources, lends this methods to analysis of the carbon budget to the finest detail as desired and as possible with available data.

\section{Concluding Remarks} \label{sect:conclusion}

We propose using the method described in Section~2c(\ref{sect:MLO-range}) and Section~2c(\ref{sect:MLO-FF}) in all further analysis of the carbon budget. Any collection of sinks or sources can be analyzed with this method if their isotopic contents and net rates of emission or sequestration are known.  Through the process of incorporating sources and sinks described here, the predominant processes affecting the Earth's carbon budget can be evaluated. When all significant sources and sinks have been accounted for, the net contributions from the biotic and abiotic pools will be zero. 

There are caveats to this type of analysis. Limitations in the available data and the algorithm mean that, in some cases, ambiguities can arise where particular differing combinations of carbon ratios and flows can lead to similar conclusions. However, under physically and biologically reasonable choices of parameter values, situations which minimize flow between the pools and the atmosphere also preserve the direction of the flow, i.e. sources remain sources and sinks remains sinks.

Finally, it would be interesting to adapt the approach which allows for variable isotope ratios to more complex model situations.  Allowing carbon pools to assume a range of values instead of the usual \emph{a priori} fixed value could shed light on more subtle mechanisms affecting the carbon budget.



\section*{Data Access} {\textsc{Matlab} files for the data, algorithms, and figures are available online.}

\section*{Author Contributions}

{All authors have made significant contributions to the paper.  All authors contributed to the editing of the manuscript. AN made all computations and wrote the first draft of the paper. CL provided necessary ecological background, framework, and motivation, and refactored the program which implements the data analysis to the form that appears in Appendix B. AN and RM contributed to the development of the mathematical method to account for known and unknown sources and sinks of carbon. CL and RM acquired funds for the project. }

{The authors declare no competing interests.}

\section*{Funding}
{The authors acknowledge the support of the Mathematics and Climate Research Network (NSF Grants DMS-0940366 and DMS-0940363), the Mathematical Sciences Postdoctoral Research Fellowship  (Award Number DMS-1902887), the Institute on the Environment at the University of Minnesota, and an Interdisciplinary Doctoral Fellowship from the Graduate School at the University of Minnesota.}

\section*{Acknowledgements}

{The authors would like to thank Eville Gorham for his considerable guidance in the ecology motivating the methods described here.  The authors would also like to thank the University of Minnesota Ecology Theory Group for workshopping the details of the methods with the authors.}


\appendix

\section{Supplementary Figures}
Below are the results of different numerical test cases for different ranges for abiotic (orange, solid or dot-dashed) and biotic (green, solid) $\delta^{13}$C signatures. We organize these test cases in $3\times3$ grids for both the current climate data from Section~\ref{sect:MLO} and paleoclimate data from Section~\ref{sect:Paleo}. We show only the cumulative curves here for brevity. Interested readers may contact the corresponding author for the rate curves if interested.

For figures where the isotope ratio may vary between fixed bounds, the top row of each figure has a biotic carbon $\delta^{13}$C of the full range of $-33 \permil$ to $-11 \permil$, representing much of the full range of signatures present in biological material \cite{Farquhar,Rounick1986}. The middle row depicts cases where the biotic carbon $\delta^{13}$C ranges from $-16 \permil$ to $-11 \permil$, the typical range for C4 photosynthesis \cite{Farquhar,Rounick1986}, and those in the bottom row range from $-33\permil$ to $-24 \permil$, the typical range for C3 photosynthesis \cite{Farquhar,Rounick1986}.   When the ratios are fixed, we take the midpoint of each of these intervals.

Since the dissolved carbon in the surface and deep ocean have isotopic signatures close to the standard ($0\permil$) \cite{Kohler, Sigman, Ciais1995a, Francey1995}, the isotope ratio bounds for the abiotic pools define ranges close to the standard. The left column of each figure depicts abiotic carbon $\delta^{13}$C restricted to the range of $-10 \permil$ to $0 \permil$. The middle column shows a range of $-6 \permil$ to $6 \permil$, and those in the right column have an abiotic carbon $\delta^{13}$C range of $0 \permil$ to $10 \permil$.  When the ratios are fixed, we take the midpoint of each of these intervals.

\begin{figure}
\begin{framed}
\begin{center}
\includegraphics[width=.7\textwidth]{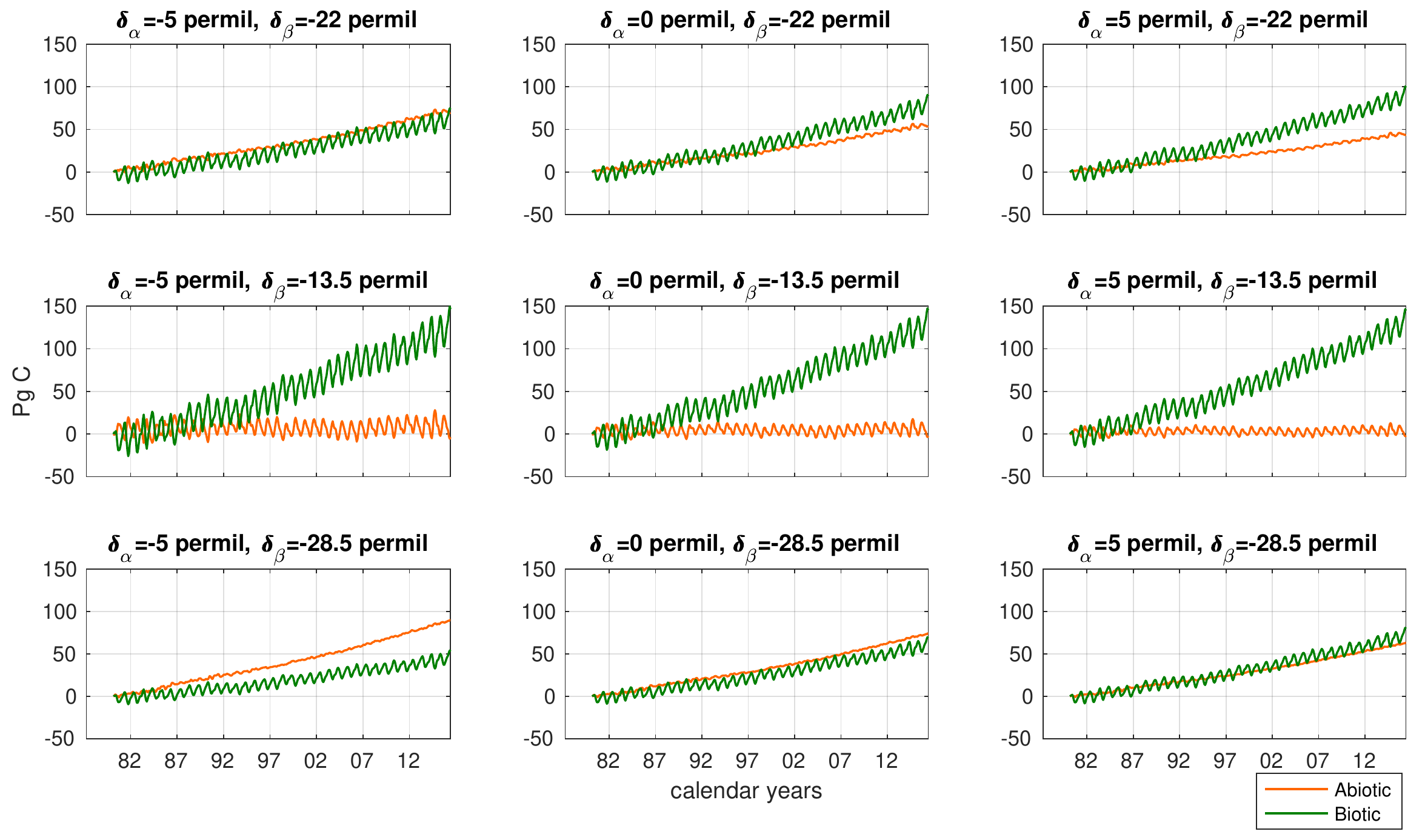}
\caption{\textbf{Mauna Loa Data with Fixed Isotope Ratios:} Cumulative contribution from the biotic and abiotic boxes.}
\label{fig:fixedMLO}
\end{center}
\end{framed}
\end{figure}

\begin{figure}
\begin{framed}
\begin{center}
\includegraphics[width=.9\textwidth]{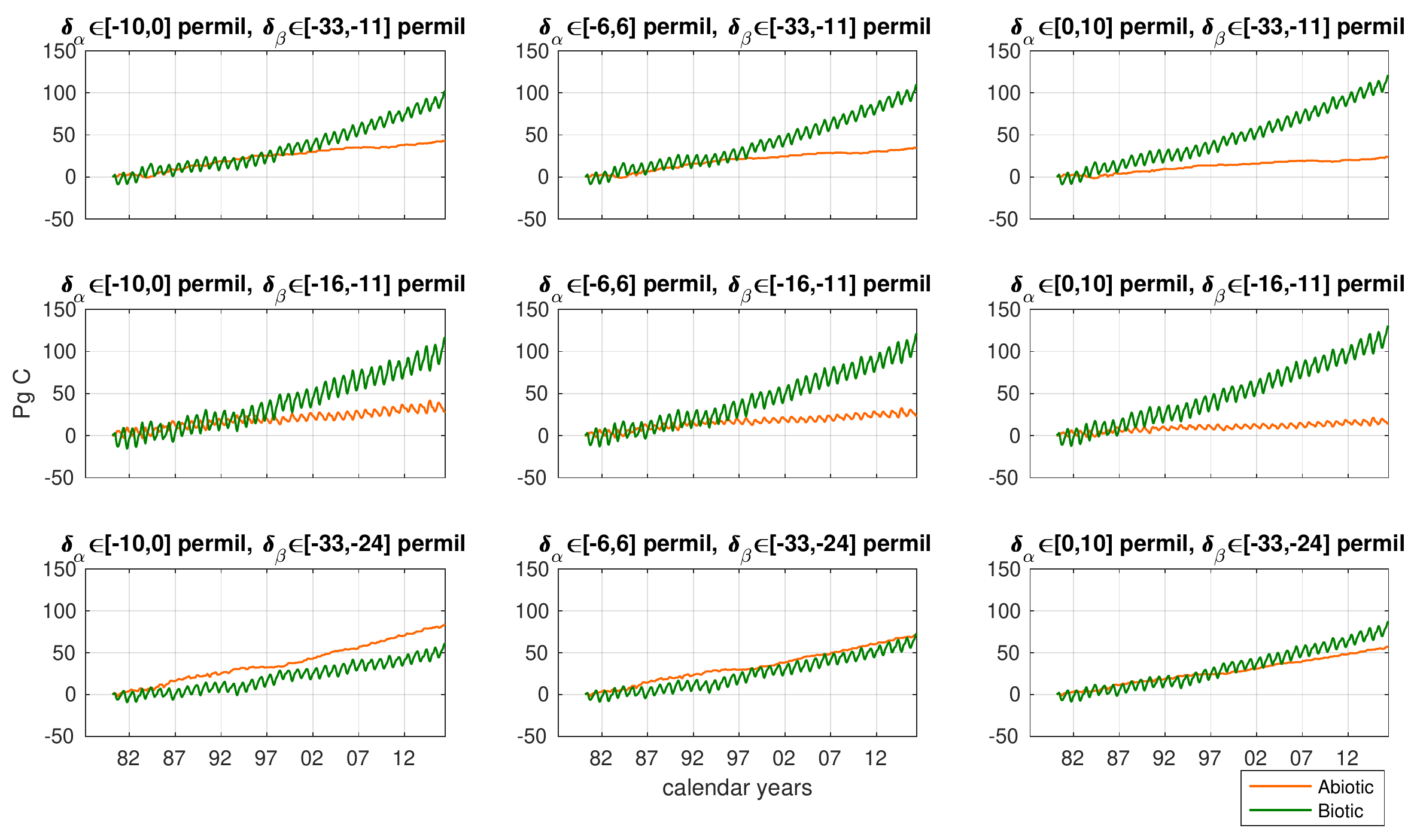}
\caption{\textbf{Mauna Loa Data with Variable Isotope Ratios:} Cumulative contribution from the biotic and abiotic boxes.}
\label{fig:rangeMLO}
\end{center}
\end{framed}
\end{figure}

\begin{figure}
\begin{framed}
\begin{center}
\includegraphics[width=.9\textwidth]{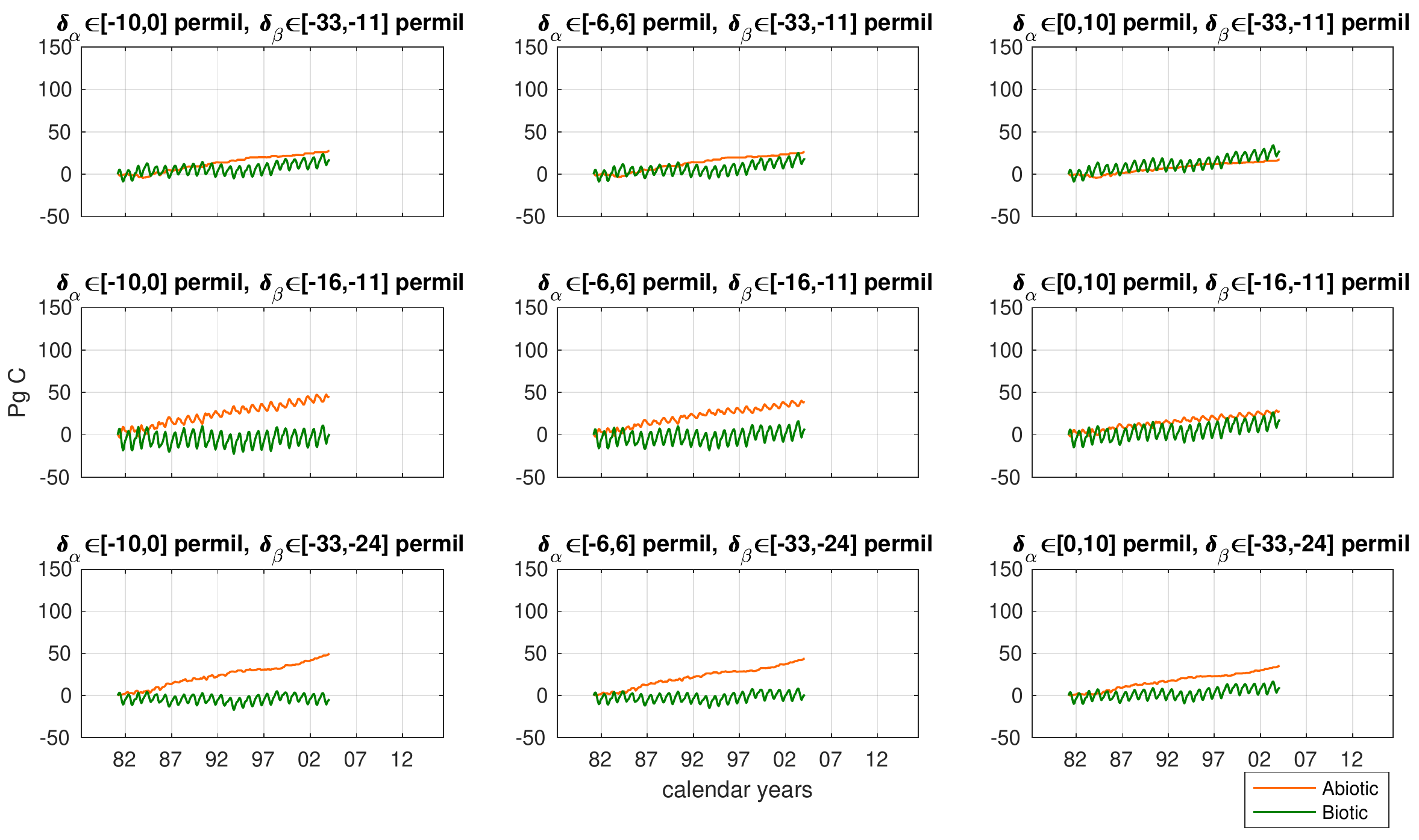}
\caption{\textbf{Mauna Loa and Fossil Fuel Data with Variable Isotope Ratios:} Cumulative contribution from the biotic and abiotic boxes.}
\label{fig:sourceMLO}
\end{center}
\end{framed}
\end{figure}

\begin{figure}
\begin{framed}
\begin{center}
\includegraphics[width=.9\textwidth]{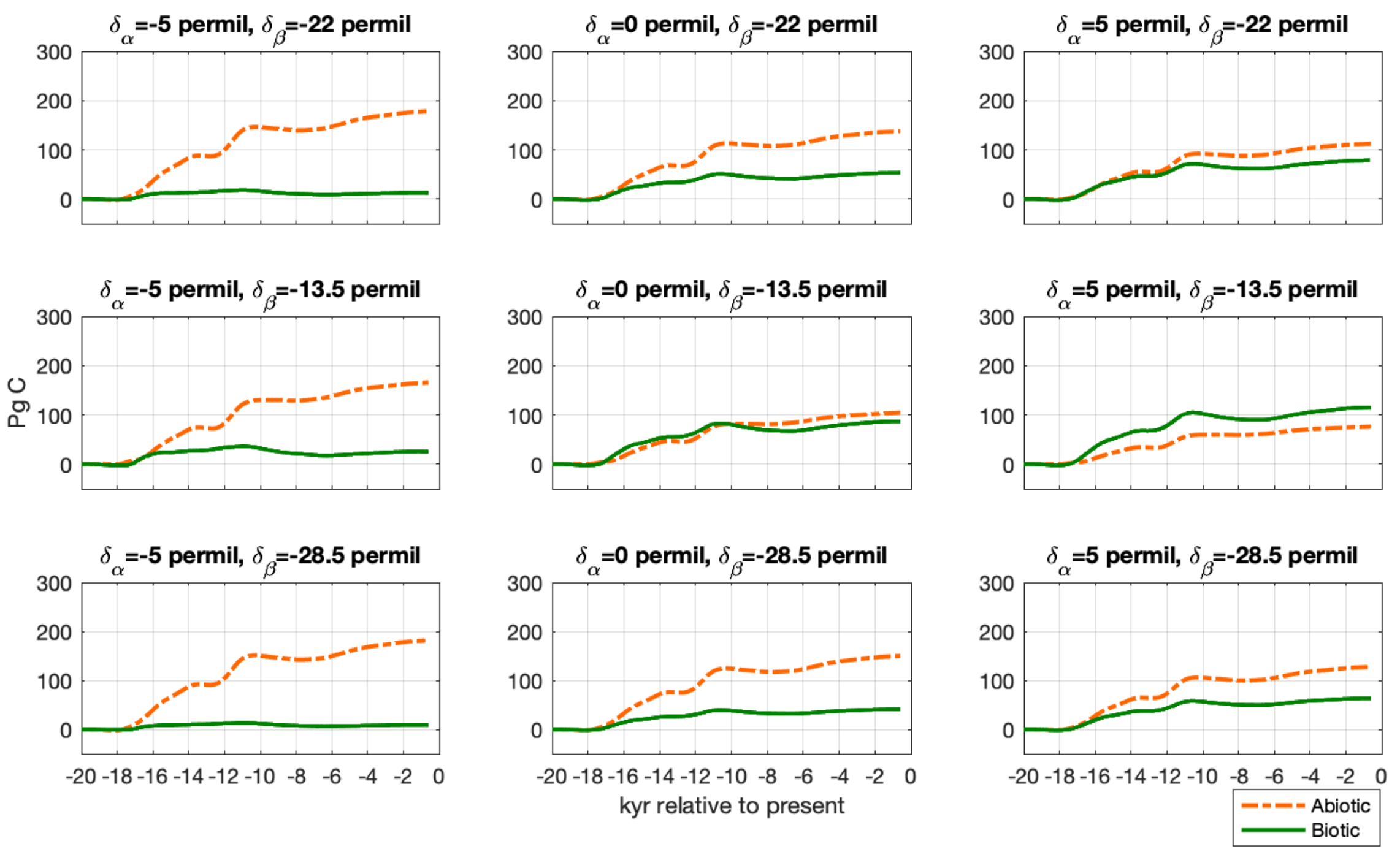}
\caption{\textbf{Paleocarbon Data with Fixed Isotope Ratios:} Cumulative contribution from the biotic and abiotic boxes.}
\label{fig:fixedPaleo}
\end{center}
\end{framed}
\end{figure}

\begin{figure}
\begin{framed}
\begin{center}
\includegraphics[width=.9\textwidth]{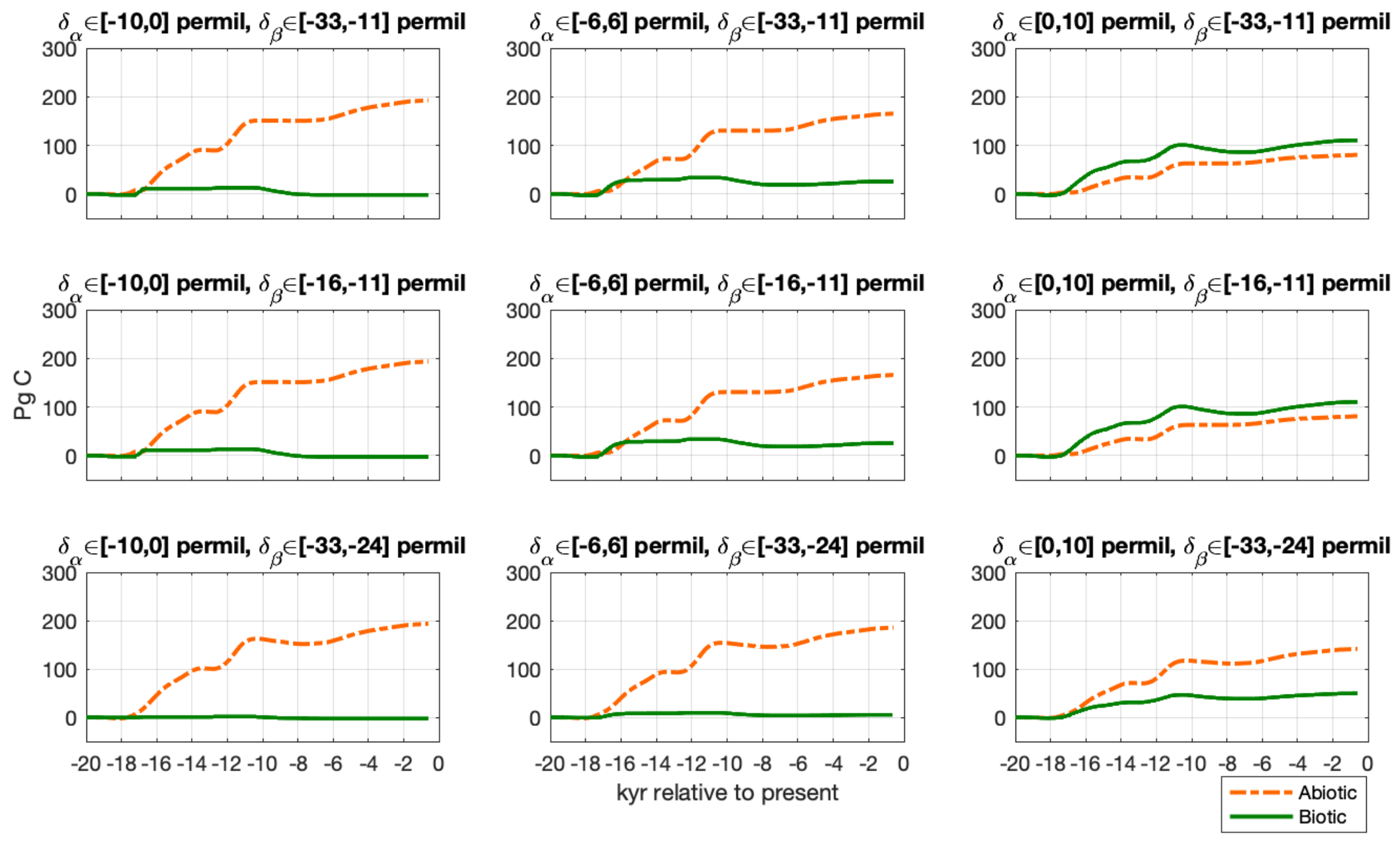}
\caption{\textbf{Paleocarbon Data with Variable Isotope Ratios:} Cumulative contribution from the biotic and abiotic boxes.}
\label{fig:rangePaleo}
\end{center}
\end{framed}
\end{figure}

\begin{figure}
\begin{framed}
\begin{center}
\includegraphics[width=.9\textwidth]{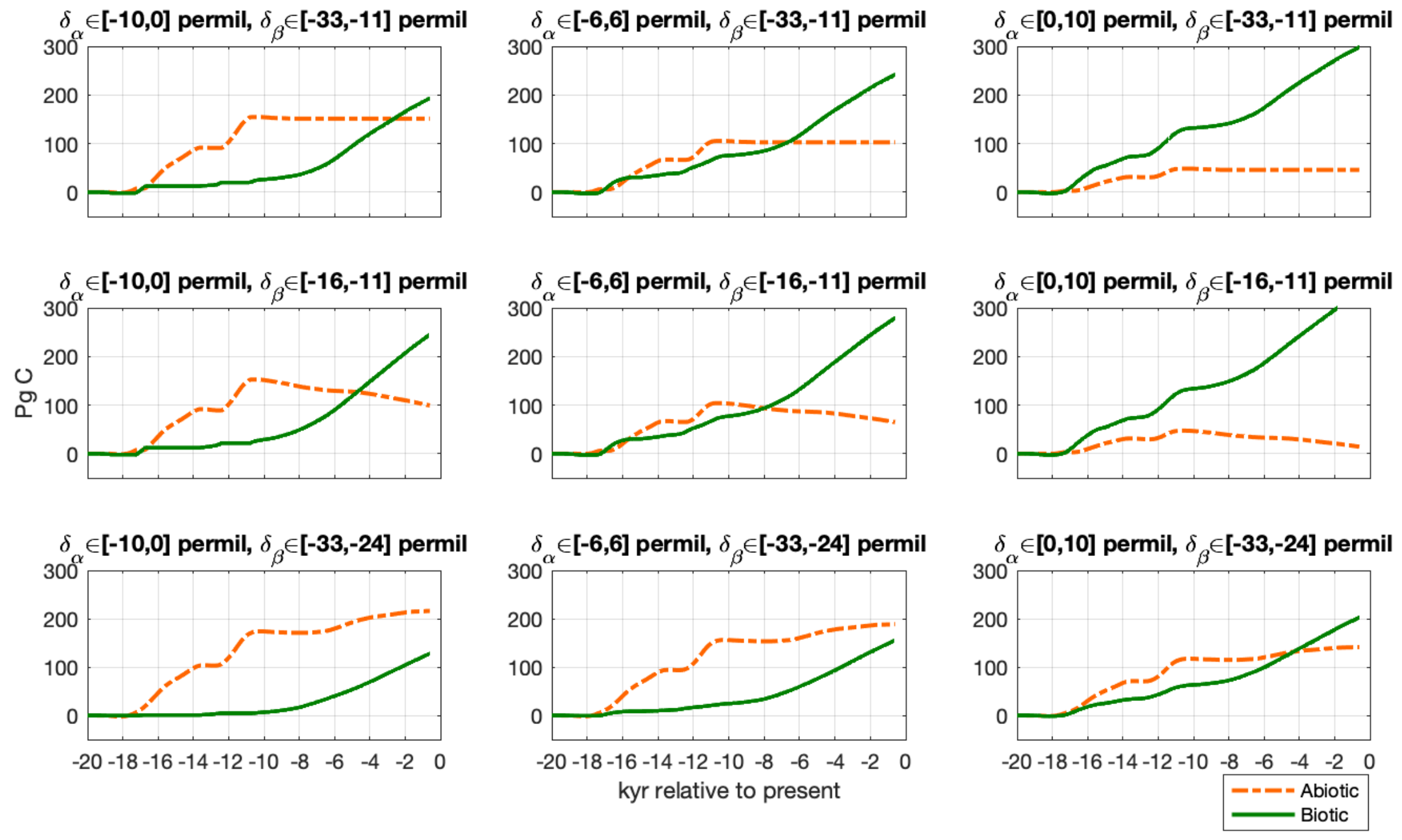}
\caption{\textbf{Paleocarbon and Peatland Data with Variable Isotope Ratios:} Cumulative contribution from the biotic and abiotic boxes.}
\label{fig:sinkPaleo}
\end{center}
\end{framed}
\end{figure}

\section{Psuedocode} \label{Sect:Pseudocode}
To allow complete evaluation of the algorithm, and to support modifications, we
present the minimization algorithm from Section \ref{sect:MLO-range} in a form of pseudo-code inspired by and simplified from
the programming languages C, R, and Python. In addition, code lines begin with a
vertical bar ($|$), as do comments on the right, in what we call ``document
format.''

The pseudo-code is two-dimensional, as in the
language Python, so that indentation completely defines the nested structure
without use of bracketing characters such as `$\{$' and `$\}$'. Statement
separators such as semicolon (`\texttt{;}') are likewise unused. Variables and
function names are italicized and flow control and reserved words are bolded.

Computations follow a left-to-right, top-to-bottom flow. Thus the assignment
operator is represented as `$\rightarrow$', similar to possibilities in R,
and also as represented in von Neumann's early computer programs.
A compound assignment such as `$x \!\rightarrow\!y\!\rightarrow\!z$' first transfers $x$ to $y$, then
transfers $y$ to $z$. Upon completion all three thus carry the same value.
Flow control with if--then--else and looping are similar to other languages.

Mathematical symbols may appear directly in the pseudocode for clarity, rather 
than being spelled out. For example, `$\alpha_1$' may be used rather than
`alpha1'.
Case matters, so that $a$ is different than $A$.
Simple functions are defined by the form
`$H(\hbox{parameters})\equiv\hbox{expression}$'. For example, one can write
`$\mu(x,y)\equiv (x+y)/2$'.
Variables not defined start out at 0.
Operator precedence is that of C.

A package is provided in {\sc {\sc Matlab}} that gives all of the code used in this manuscript as well as the necessary code to generate the figures. Equivalent versions of this algorithm translated into operational C are available from the authors upon request.

\def\g{\hskip.5em}

The program begins with the definition of three functions and initialization of
bounds of the carbon pools.

\begin{figure}
\begin{framed}
\caption{Pseudocode for the Minimization Algorithm}
\vspace{5mm}
\small
On entry to the program, input consists of a data file in order of increasing
time. The input data have three columns.

\begin{itemize}
\item[$n$:] Sequence number representing a variable-length time step,
             for reference.

\item[$A$:] Flow of carbon 12 during the time step, in petagrams. Positive is
             flow from carbon pools to the atmosphere, negative is the reverse.

\item[$B$:] Flow of carbon 13 during the time step, in petagrams. Positive is
             flow from carbon pools to the atmosphere, negative is the reverse.
\end{itemize}

On exit from the program, output consists of a data file in order of increasing
time. The output data have seven columns.

\begin{itemize}
\item[$n$:] A sequence number representing the time step,
             matching the first column in the input file.

\item[$A$:] Flow of carbon 12 during the time step,
             matching the corresponding column in the input file.

\item[$B$:] Flow of carbon 13 during the time step,
             matching the corresponding column in the input file.

\item[$x$:] Total carbon entering or leaving the set of abiotic
             carbon pools during the time step, in petagrams.

\item[$y$:] Total carbon entering or leaving the set of biotic
             carbon pools during the time step, in petagrams.

\item[$\alpha$:]
             Ratio of carbon 13 to total carbon in the flow represented by $x$
             in this time step.

\item[$\beta$:]
             Ratio of carbon 12 to total carbon in the flow represented by $y$
             in this time step.
\end{itemize}

\vspace{5mm}

\includegraphics[width=.99\textwidth]{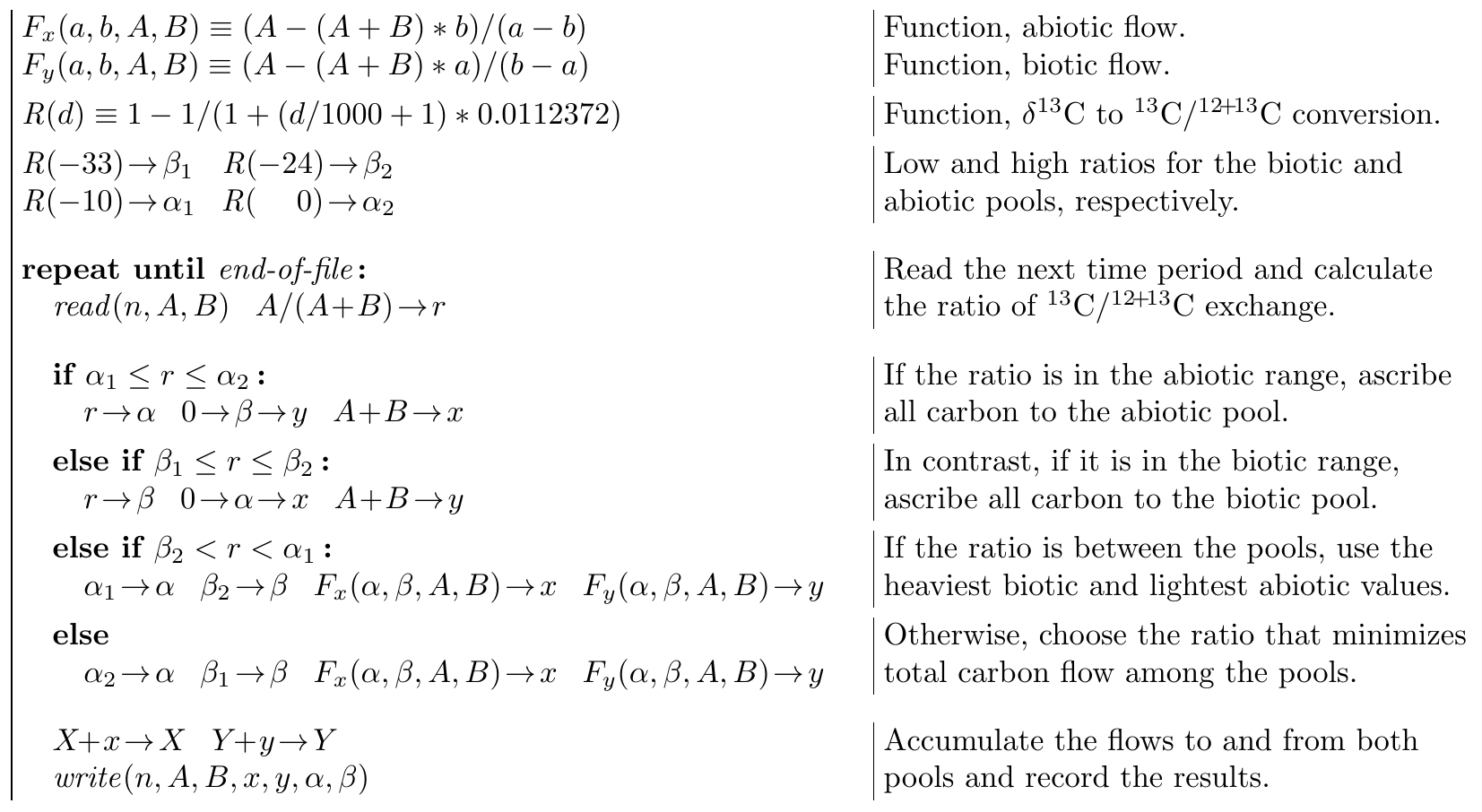}
\end{framed}
\end{figure}


\begin{thebibliography}{1}

%
\bibitem{Andersson} Andersson, Rina Argelia, et al. ``Elemental and isotopic carbon and nitrogen records of organic matter accumulation in a Holocene permafrost peat sequence in the East European Russian Arctic.'' \emph{Journal of Quaternary Science} 27.6 (2012): 545-552.

%
\bibitem{Aucour} Aucour, Anne-Marie, Raymonde Bonnefille, and Claude Hillaire-Marcel. ``Sources and accumulation rates of organic carbon in an equatorial peat bog (Burundi, East Africa) during the Holocene: carbon isotope constraints.'' \emph{Palaeogeography, Palaeoclimatology, Palaeoecology} 150.3-4 (1999): 179-189.

%
\bibitem{Blasing} Blasing, T.J., C.T. Broniak, and G. Marland, 2004. ``Estimates of monthly carbon dioxide emissions and associated $\delC$ values from fossil-fuel consumption in the U.S.A.'' \emph{Trends: A Compendium of Data on Global Change}, Carbon Dioxide Information Analysis Center, Oak Ridge National Laboratory, U.S. Department of Energy, Oak Ridge, TN, U.S.A. doi: 10.3334/CDIAC/ffe.001

%
\bibitem{Berner1991} Berner, Robert. ``A model for atmospheric CO2 over Phanerozoic time.'' \emph{American Journal of Science,} 291.4. (1991): 339-376.

%
\bibitem{Berner2001} Berner, Robert, and Zavareth Kothavala. ``GEOCARB III: A revised model of Atmospheric CO2 over Phanerozoic time.'' \emph{American Journal of Science,} 301.2. (2001): 182-204.

%
\bibitem{Berner2003} Berner, Robert. ``The long-term carbon cycle, fossil fuels and atmospheric composition.'' \emph{Nature,} 426. (2003): 323-326.

%
\bibitem{Ciais1995a} Ciais, P., P. P. Tans, J. W. C. White, M. Trolier, R. J. Francey, J. Berry, D. R. Randall, P. J. Sellers, J. G. Collatz, and D. S. Schimel. ``Partitioning of ocean and land uptake of CO2 as inferred by $\delta^{13}$C measurements from the NOAA Climate Monitoring and Diagnostics Laboratory Global Air Sampling Network.'' \emph{Journal of Geophysical Research,} 100. (1995): 5051-5070.

%
\bibitem{Cristea} Cristea, Gabriela, et al. "Carbon isotope composition as indicator for climatic changes during the middle and late Holocene in a peat bog from Maramures Mountains (Romania)." \emph{The Holocene} 24.1 (2014): 15-23.

%
\bibitem{Farquhar} Farquhar, G. D., J. R. Ehleringer, and K. T. Hubick. ``Carbon isotope discrimination and photosynthesis,'' \emph{Annu. Rev. Plant Physiol. Plant Mol. Biol.} 40. (1989): 503-537.

%
\bibitem{Francey1995} Francey, R., P. Tans, C. Allison, I. Enting, J. W. C. White, M. Trolier. ``Changes in oceanic terrestrial carbon uptake since 1982.'' \emph{Nature} 373. (1995): 326-330.

%
\bibitem{VPDB}  Gonfiantini, R., W. Stichler, K. Rozanski.  ``Standards and intercomparison materials distributed by the International Atomic Energy Agency for stable isotope measurements.''  \emph{Reference and intercomparison materials for stable isotopes of light elements.} IAEA-TECDOC-825 (1993):13-29.

%
\bibitem{Gorham91} Gorham, Eville. "Northern peatlands: role in the carbon cycle and probable responses to climatic warming." \emph{Ecological applications} 1.2 (1991): 182-195.

%
\bibitem{Gorham07} Gorham, Eville, et al. "Temporal and spatial aspects of peatland initiation following deglaciation in North America." \emph{Quaternary Science Reviews} 26.3-4 (2007): 300-311.

%
\bibitem{Gorham12} Gorham, Eville, et al. "Long-term carbon sequestration in North American peatlands." \emph{Quaternary Science Reviews} 58 (2012): 77-82.

%
\bibitem{Kaplan} Kaplan, Jed O., et al. "Holocene carbon emissions as a result of anthropogenic land cover change." \emph{The Holocene} 21.5 (2011): 775-791.

%
\bibitem{Keeling} Keeling, Charles D., et al. "Atmospheric CO 2 and 13 CO 2 exchange with the terrestrial biosphere and oceans from 1978 to 2000: Observations and carbon cycle implications." \emph{A history of atmospheric CO2 and its effects on plants, animals, and ecosystems.} Springer, New York, NY, 2005. 83-113.

%
\bibitem{Kohler} K\"ohler, P., et al. "Quantitative interpretation of atmospheric carbon records over the last glacial termination.'' \emph{Global Biogeochemical Cycles} 19. (2005): GB4020.

%
\bibitem{Kruger} Kr\"uger, J. P., et al. "Biogeochemical indicators of peatland degradation-a case study of a temperate bog in northern Germany." \emph{Biogeosciences} 12 (2015): 2861-2871. 

%
\bibitem{Mills2019} Mills, Benjamin, et al. "Modelling the long-term carbon cycle, atmospheric CO2, and Earth surface temperature from late Neoproterozoic to present day." \emph{Gonwana Research} 67 (2019): 172-186. 

%
\bibitem{Monnin} Monnin, Eric. [DATASET] EPICA Dome C high resolution carbon dioxide concentrations. PANGAEA, https://doi.org/10.1594/PANGAEA.472488

%
\bibitem{Reinsch} Reinsch, C. ``Smoothing by Spline Functions.'' \emph{Numerische Mathematik,} 10. (1967): 177-183.

%
\bibitem{Rounick1986} Roundick, J., and M. Winderbourn. ``Carbon Isotopes and Carbon Flow in Ecosystems.'' \emph{BioScience,} 36.3. (1986): 171-177. 

%
\bibitem{Ruddiman} Ruddiman, W. ``The Anthropocene,'' \emph{Annu. Rev. Earth Palnet. Sci.} 41. (2013): 45-68.

%
\bibitem{Schmitt} Schmitt, Jochen, et al. "Carbon isotope constraints on the deglacial CO2 rise from ice cores." \emph{Science} 336.6082 (2012): 711-714.

%
\bibitem{Shields2017} Shields, Craham, and Benjamin Mills. "Tectonic controls on the long-term carbon isotope mass balance." \emph{PNAS} 114.17 (2017): 4318-4323. doi: 10.1073/pnas.1614506114

%
\bibitem{Sigman} Sigman, Daniel M., Mathis P. Hain, and Gerald H. Haug. "The polar ocean and glacial cycles in atmospheric CO 2 concentration." \emph{Nature} 466.7302 (2010): 47-55.

%
\bibitem{Skrypek} Skrzypek, Grzegorz, et al. "The carbon stable isotopic composition of mosses: A record of temperature variation." \emph{Organic geochemistry} 38.10 (2007): 1770-1781.

%
\bibitem{Zhang} Zhang, Jiawu, et al. "Holocene monsoon climate documented by oxygen and carbon isotopes from lake sediments and peat bogs in China: a review and synthesis." \emph{Quaternary Science Reviews} 30.15-16 (2011): 1973-1987.

%
\bibitem{Zhong} Zhong, Wei, et al. "Carbon isotope evidence of last glacial climate variations in the tropical NW Leizhou Peninsula, South China." \emph{Boreas} 41.1 (2012): 102-112.

\end{thebibliography}
\end{document}